\documentclass[aps,prd,groupedaddress,preprint,eqsecnum,nofootinbib,showpacs]{revtex4}
\usepackage{graphicx,epsf,amssymb,amsbsy,amsfonts,amssymb,amsmath}
\usepackage[dvips]{color}
\newif\ifdraft
\drafttrue
\newif\ifpreprint
\preprinttrue

\def\sect#1{section~{\ref{#1}}}
\def\fig#1{fig.~{\ref{#1}}}

\def\tab#1{table~{\ref{#1}}}

\def\spa#1.#2{\left\langle#1\,#2\right\rangle}
\def\spb#1.#2{\left[#1\,#2\right]}

\def\Tr{\, {\rm Tr}}

\def\eps{\epsilon}

\def\nn{\nonumber}

\def\eqn#1{eq.~(\ref{#1})}

\def\NeqFour{{{\cal N}=4}}

\def\NeqEight{{{\cal N}=8}}

\def\f#1{\tilde f^{#1}}
\def\be{\begin{equation}}
\def\ee{\end{equation}}
\def\bea{\begin{eqnarray}}
\def\eea{\end{eqnarray}}
\def\ba{\begin{eqnarray}}
\def\ea{\end{eqnarray}}

\def\MHVbar{$\overline{\hbox{MHV}}$}

\def\Frac#1#2{{\textstyle \frac{#1}{#2}}}
\def\I{{\cal I}}

\newbox\charbox
\newbox\slabox
\def\s#1{{   
 \setbox\charbox=\hbox{$#1$}
 \setbox\slabox=\hbox{$/$}
 \dimen\charbox=\ht\slabox
 \advance\dimen\charbox by -\dp\slabox
 \advance\dimen\charbox by -\ht\charbox
 \advance\dimen\charbox by \dp\charbox
 \divide\dimen\charbox by 2
 \raise-\dimen\charbox\hbox to \wd\charbox{\hss/\hss}
 \llap{$#1$} }}

\def\f{\tilde f}

\begin{document}
\hfuzz 20pt

\ifpreprint
SU-ITP-11/32  $\null\hskip0.1cm\null$ \hfill
Saclay--IPhT--T11/152
\fi

\title{Five-Point Amplitudes in $\NeqFour$ Super-Yang-Mills Theory\\ and $\NeqEight$ Supergravity}

\author{John~Joseph~M.~Carrasco${}^a$ and Henrik~Johansson${}^b$}

\affiliation{
${}^a$Stanford Institute for Theoretical Physics and Department of Physics, Stanford University,
             Stanford, CA 94305-4060, USA \\
${}^b$Institut de Physique Th\'eorique, CEA--Saclay,
  F--91191 Gif-sur-Yvette cedex, France\\
}

\email{jjmc@stanford.edu, henrik.johansson@cea.fr}

\date{June, 2011}

\begin{abstract}

We present the complete integrands of five-point superamplitudes in $\NeqFour$ super-Yang-Mills theory and $\NeqEight$ supergravity, at one and two loops, for four-dimensional external states and $D$-dimensional internal kinematics.  For $\NeqFour$ super-Yang-Mills theory we give the amplitudes for general gauge group -- including all nonplanar contributions.  The results are constructed using integral diagrams that manifestly satisfy the conjectured duality between color and kinematics, providing additional nontrivial evidence in favor of the duality for multipoint and multiloop amplitudes. We determine the ultraviolet poles by integrating the amplitudes in the dimensions where logarithmic divergences first occur. We introduce new kinematic prefactors which offer a convenient decomposition of the external state structure of the nonplanar five-point amplitudes in the maximally supersymmetric theories to all loop orders. 
\end{abstract}

\pacs{04.65.+e, 11.15.Bt, 11.30.Pb, 11.55.Bq \hspace{1cm}}

\maketitle


\section{Introduction}
\label{Introduction}

The computation of scattering amplitudes has proven a remarkably rewarding theoretical problem, exposing previously unknown symmetries and structures of well-studied field theories. Spectacular examples of this can be found in the maximally supersymmetric $\NeqFour$ super-Yang-Mills theory (sYM)~\cite{NeqFourDefinition}, where structures such as dual conformal symmetry~\cite{MagicIdentities, DualSuperconformalSymmetry}, twistor string theory~\cite{WittenTwistorString} and Grassmannians~\cite{Grassmannians} have emerged out of the study of on-shell amplitudes.

A structure of particular relevance to nonplanar gauge theory and to gravity, is the duality between kinematic and color constituents of amplitudes, uncovered at tree-level~\cite{BCJ}  and conjectured to extend to all loop orders~\cite{BCJLoop} by Bern and the current authors.  The duality has the property of interlocking the various kinematic diagrams of generic gauge theories into a very rigid system, effectively minimizing the independent information needed to specify an amplitude. At tree level, this has been used to construct an $(n-3)!$ basis for color-ordered $n$-point amplitudes~\cite{BCJ}, which has since been proven in both string and field theory~\cite{Monodromy,amplituderelationProof}.  At the mulitloop level, it has been used to specify the complete (planar and nonplanar) integrands for special four-point amplitudes using one or more ``master graphs''~\cite{BCJLoop,ck4l}.

A second property of the duality is the remarkably simple structure imposed on gravity amplitudes. Once gauge theory amplitudes are arranged such that the kinematic factors are on equal footing with color factors, then gravity amplitudes are obtained through simple double copies of the gauge theory kinematic factors~\cite{BCJ, BCJLoop}. In ref.~\cite{Square} this was given a detailed proof valid for tree-level amplitudes, in the case of $\NeqEight$ supergravity~\cite{CremmerJulia} as a double copy of $\NeqFour$ sYM, and in the case where Einstein gravity amplitudes are acquired through double copies of pure Yang-Mills amplitudes.  At tree level the double-copy structure is closely related to the Kawai-Lewellen-Tye relations~\cite{KLT} between closed and open string amplitudes.  Beyond potentially clarifying the inner structure of gravity, the double-copy property offers a way to circumvent the usually cumbersome computations of integrands of loop-level gravity scattering amplitudes.  

The duality between kinematics and color is conjectured to be valid for generic Yang-Mills and gravity theories in any dimension and to any loop order and multiplicity.  At tree level, strong supporting evidence exists~\cite{BCJ,Square,Monodromy,amplituderelationProof,Bjerrum2, Tye,KiermaierTalk, BBDSVsolution, Feng,ConnellKinematicAlgebra, StiebergerNumerators}; see also ref.~\cite{BCJOther} for various applications. As for loop level, four-point calculations in the $\NeqFour$ sYM and $\NeqEight$ supergravity theories have shown that duality-satisfying amplitude representations can be found through four loops~\cite{BCJLoop,ck4l}, and also for the two-loop four-point identical-helicity amplitude in QCD~\cite{BCJLoop}.

The natural implication of the observed duality is that the kinematic structures of both gauge theories and gravity theories are elements of some hereto possibly unknown Lie algebras. In a recent paper by Bern and Dennen~\cite{BernDennenTrace} this was assumed in making the first steps towards a trace representation of the algebra. Even more recently Monteiro and O'Connell~\cite{ConnellKinematicAlgebra} identified a certain infinite-dimensional area-preserving diffeomorphism algebra in the self-dual sector of Yang-Mills theory as being responsible for the duality, at least for the case of maximally-helicity-violating (MHV) tree amplitudes in four dimensions. For non-MHV amplitudes, or higher-dimensional amplitudes the algebra is not yet known. A step in this direction was taken in~\cite{Square}, where the first terms in a cubic Yang-Mills  Lagrangian that obeyed the duality were worked out. Knowing the full form of such a Lagrangian would be equivalent to knowing the structure constants of the kinematic algebra, at least at the level of tree amplitudes.  

In this paper we strengthen the evidence supporting the duality by explicitly computing several duality-satisfying five-point amplitudes in $\NeqFour$~sYM and, consequently, in $\NeqEight$~supergravity. First we work out a representation of the four-dimensional duality-satisfying one-loop five-point amplitude in detail. This amplitude will turn out to have an interesting and compact structure that encodes the external state dependence.  Based on this structure the pattern for higher-loop five-point amplitudes will become clear, resulting in a proposed ansatz for the duality-satisfying amplitude to arbitrarily loop order at five points. This ansatz is parametrized by rational coefficients, which require further explicit calculations in order to be determined. We confirm the ansatz through three loops by computing the amplitudes at each loop order. The one- and two-loop results are included in this paper, and the three-loop results will be given in an accompanying paper~\cite{threeloopfivept}. 

The planar five-point $\NeqFour$ sYM amplitudes have previously been computed through three loops~\cite{MHVoneloop, KorchemskyOneLoop,FivePtBDS,  CachazoLeadingSingularityAndCalcs, Spradlin3loop}. Beyond one loop, nonplanar amplitudes have not been worked out (other than for particular four-point amplitudes~\cite{BRY, compact3, Neq44np}), so the results presented here are novel.   The four-dimensional one-loop five-point amplitude of $\NeqEight$ supergravity is previously known~\cite{MHVOneLoopGravity}. However, the form of the amplitude presented here is more general as it is valid for all values of the dimension of the internal momenta, or dimensional regularization parameter, $D$ (in \cite{MHVOneLoopGravity} the closely related all-plus-helicity amplitude was given for general $D$). Similarly, the two-loop five-point $\NeqEight$ supergravity amplitude is a new result. All amplitudes presented in this paper will be valid for general internal $D$, and the gauge theory amplitudes will be given for general gauge theory group $G$. 

The study of potential ultraviolet (UV) divergences and counterterms of $\NeqEight$ supergravity is an area of active research, see {\it e.g.}~\cite{UVreview, StringsE7andSusy,BeisertCountertermsandElvangReview,RecentDualityStuff,HoweStelle}. It was proposed in \cite{issupergravityfinite} that this theory may be finite to all loop orders, contrary to common expectations.  Spectacular ultraviolet cancellations were subsequently observed in direct calculations of the three- and four-loop four-point amplitudes~\cite{superfiniteness,compact3,Gravityfourloops}. Recently, counterterm analysis and other methods have ruled out divergences through at least six loops in four dimensions~\cite{BeisertCountertermsandElvangReview,Bjornsson,BossardVanishingVolume,Kallosh7loops}. In this paper we supplement these results with the more modest five-point one- and two-loop ultraviolet studies. We explicitly integrate the newly obtained amplitudes in the lowest (critical) dimensions where they develop a ultraviolet divergence, namely $D=8$ at one loop, and $D=7$ at two loops, for both $\NeqFour$ sYM and $\NeqEight$ supergravity. We find that the general form of the divergences of the two theories and two amplitudes agree with the observed divergences of the corresponding four-point amplitudes~\cite{GSB,MarcusSagnottii,BernDixonUV}. Thus, in doing so, we verify the expected UV behavior of these theories~\cite{BernDixonUV,HoweStelle,BRY} at five points through two loops.  
 
For the construction of the amplitudes we implicitly make use of the unitarity method~\cite{UnitarityMethod} and generalized unitarity~\cite{GeneralizedUnitarity,GeneralizedUnitarity2,BCF}, which we will not discuss in any detail. Recent reviews on these very topics can be found in refs.~\cite{ZviYutinReview, DixonReview, JJMCHJreview, BrittoReview}.
 
The organization of the paper is as follows. In \sect{Review}, we set up notation and review the organization of amplitudes that satisfy the duality between color and kinematics, as well as  preview the general multiloop five-point structure.   In \sect{OneLoopSection}, we construct the duality-satisfying one-loop amplitudes, and compute their corresponding ultraviolet divergences.  Similarly, in \sect{twoloopsection} we construct the duality-satisfying two-loop amplitudes, and compute their corresponding ultraviolet divergences. Finally, in \sect{Conclusions} we present our conclusions. 

\section{Review and method}
\label{Review}
\subsection{Cubic and duality-satisfying representations}

The conjectured  duality between color and kinematic factors~\cite{BCJ, BCJLoop} relies upon a representation of gauge theory amplitudes using graphs with only cubic vertices, see {\it e.g.} reviews~\cite{JJMCHJreview,StructureReview}. For five-point amplitudes in the adjoint representation we have
\be
{\cal A}_5^{(L)}= i^L g^{3+2L}\sum_{i\in\Gamma_3} \, \int \frac{d^{LD}p}{(2\pi)^{LD}} \frac{1}{S_i} \, \frac{N_i C_i}{l_{i_1}^2l_{i_2}^2l_{i_3}^2\cdots l_{i_{m}}^2}\,.
\label{LoopBCJform}
\ee
where $d^{LD}p=\prod_{j=1}^{L}d^Dp_j$ is the usual integral measure of $L$ independent $D$-dimensional loop momenta $p_j^\mu$, and $\Gamma_3$ is the set of all cubic $L$-loop five-leg graphs, counting all relabeling of external legs. Corresponding to each internal line (edge) of the $i^{\rm th}$ graph we associate a propagator $1/l_{i_l}^2$, which is a function of the independent internal and external momenta, $p_j$ and $k_j$, respectively.  The local numerator functions, here only schematically indicated as $N_i$, include information about the kinematics and states. The color factors $C_i$ contain the information of the gauge-group structure, and are given by multiplication of the structure constants $\f^{abc}= i \sqrt{2} f^{abc} = \Tr([T^a, T^b] T^c)\,$, with Hermitian generators $T^a$ normalized via $\Tr(T^a T^b) = \delta^{ab}$. The symmetry factors $S_i$ are the same as those obtained in, say, scalar $\phi^3$ theory.

The duality between color and kinematics is satisfied in amplitude representations (\ref{LoopBCJform}) where the kinematic numerators $N_i$ obey the same general algebraic relations as the color factors $C_i$.  Specifically, the $N_i$ obey Jacobi relations and have antisymmetric behavior analogous to the color factors, schematically,
\bea
~~~~~~ N_i+N_j+N_k=0~~ &\Leftrightarrow&~~ C_i+C_j+C_k=0\,,~~~~~~~~~{\rm (Jacobi~identity)} \\
~~~~~~~~~~~~~N_{i}\rightarrow -N_{i}~ &\Leftrightarrow&~~C_{i} \rightarrow -C_{i}\, ~~~~~~~~~~~~~~~\mbox{(vertex-flip~antisymmetry)} \nn
\label{DefiningJacobi}
\eea
where the first line signifies the Jacobi identity valid for specific triplets of graphs in the amplitude, and the second line represents the action of flipping the ordering of a cubic vertex in a graph. As we will see in the following sections, at loop level it is most convenient to treat these kinematical  relations as functional equations over the internal momentum space. In addition, it can be useful to impose the self-symmetries or graph automorphisms on the $N_i$, similar to the self-symmetries obeyed by the color factors $C_i$ -- this effectively reduces the number of independent $N_i$ functions.

Once the gauge theory numerators satisfy the duality we can construct gravity numerators by taking two copies of gauge theory kinematic numerators~\cite{BCJ,BCJLoop}. Given a five-point $L$-loop gauge theory amplitude with duality-satisfying numerators $N_i$, the corresponding gravity amplitude takes the form
\be
{\cal M}_5^{(L)}= i^{L+1} \left(\frac{\kappa}{2}\right)^{3+2L}\sum_{i\in\Gamma_3} \, \int \frac{d^{LD}p}{(2\pi)^{LD}} \frac{1}{S_i} \, \frac{N_i \widetilde N_i}{l_{i_1}^2l_{i_2}^2l_{i_3}^2\cdots l_{i_{m}}^2}\,,
\label{GravBCJ}
\ee
where $\kappa$ is the gravitational coupling constant, and $\tilde{N}_i$ are a set of  kinematic numerators for the amplitude of a possibly different gauge theory (which need not explicitly satisfy the duality~\cite{BCJLoop, Square}). To construct $\NeqEight$ supergravity amplitudes, we are interested in the case where both numerators belong to the $\NeqFour$ sYM theory; hence, the two numerators are effectively identical.  However, for precise bookkeeping of the individual states we should  distinguish the $R$-symmetry indices of the two numerators. The $\NeqEight$ supergravity theory has $R$-symmetry group $SU(8)$, and the double-copy form (\ref{GravBCJ}) makes part of this manifest, namely the subgroup $SU(4)\otimes \widetilde{SU(4)}$. A convenient way to embed this into $SU(8)$ is to shift the $R$-symmetry indices of the second numerator copy $\tilde{N}_i = N_i |_{A \rightarrow A + 4}$ by a uniform offset of 4.

\subsection{Method and five-point numerator structure}
\label{GeneralStructureSection}

The procedure for finding a duality-satisfying representation for an amplitude involves several steps. Our approach is as follows:  one starts by identifying all distinct cubic graphs with $L$ loops and $m$ external legs, and writes down the linear equation system generated by the kinematic Jacobi relations and graph self-symmetries. Reducing this linear system by simple elimination of numerators eventually results in a system where very few graph numerators remain. These are the so-called ``master graph'' numerators, as they effectively encode the full amplitude. 

The kinematic Jacobi relations generates functional equations when applied to loop-diagram numerators, as the relations typically compares the numerators at different points in the internal momentum space. The functional equations are in general nontrivially entangled, but when reasonable assumptions on the form of the functions can be made such systems are readily solved.   Specifically, if locality is assumed, then the numerators are simple polynomials of a degree fixed by the engineering dimension of the numerator.  Assigning each master graph a local ansatz built out of external and internal momenta, polarizations and spinors, should be sufficient for the task. However, using formal polarization vectors usually results in overly complicated expressions that obscure the otherwise compact analytic form of the amplitude (although notable exceptions exists~\cite{StiebergerNumerators}). Similarly, such representations can impede making manifest the kinematic simplifications that occur when the external states and momenta are restricted to a fixed space-time dimension, in our case four dimensions. The way around this is to use notation that is specifically designed to simplify bookkeeping of states in that dimension. In four dimensions the spinor-helicity formalism is especially handy (see {\it e.g.}~ref.~\cite{DixonReview}). However, without using explicit polarization vectors we can no longer expect the numerators to be strictly local in external momenta.  Indeed we will see this phenomenon below.  That said, in these cases, we are able to demand locality for the internal loop momenta.

The nub of the matter is in arriving at a sufficiently general ansatz that is still compact enough to work with.  For four-point amplitudes of $\NeqFour$ sYM (at least through four loops) this is by now a well-understood problem.  It turns out that all of the external state information of a multiloop four-point amplitude can be conveniently packaged in the universal crossing-symmetric prefactor~\cite{Neq44np}
\be
{\cal K}(1,2,3,4)\equiv s_{12} s_{23} A^{\rm tree}_4 (1,2,3,4) =-i \delta^{(8)}(Q){\spb{1}.{2}\spb{3}.{4}\over \spa{1}.{2}\spa{3}.{4}}\,,
\ee
where $A_4^{\rm tree}(1,2,3,4)$ is the color-ordered $D$-dimensional four-point amplitude for any possible combination of external states (here suppressed), and $s_{ij}=(k_i+k_j)^2$. The second expression is given by plugging in the explicit $D=4$ superamplitude. The $\spa{i}.{j}$ and $\spb{i}.{j}$ are skew-symmetric products of Weyl spinors  that satisfy the property $\spa{i}.{j}\spb{j}.{i}=s_{ij}$ (see {\it e.g.}~ref.~\cite{DixonReview}). The delta function is Grassmann valued and takes as its argument the overall supermomentum of an $m$-point amplitude
\be
Q^{\alpha A}=\sum_{i=1}^{m}\lambda_i^{\alpha}\eta_i^{A}\,,
\label{gdelta}
\ee
where $\eta_i^A$ are Grassmann variables, with $A$ being a $SU(4)$ $R$-symmetry index, and $\lambda_i^{\alpha}$ is a Weyl spinor with $SU(2)$ index $\alpha$ (see {\it e.g.}~ref.~\cite{Supersums}). 

The remarkable property of ${\cal K}(1,2,3,4)$ is that it can be used to construct very compact four-point graph numerators, of the schematic form
\be
N_i \, \sim \, {\cal K}(1,2,3,4) \, \times \, ({\rm local~momentum~factor})\,,
\label{localKappa}
\ee
where the local factor is built entirely of Lorentz products of momenta, as demonstrated up to four loops~\cite{Neq44np,ck4l}.

For five-point amplitdes in $\NeqFour$ sYM, it is not clear {\it ab intitio} what the correct generalization of ${\cal K}(1,2,3,4)$ should be.  Fortunately, we will in this paper uncover a set of five-point prefactors that generalizes the behavior of  ${\cal K}(1,2,3,4)$.  The details of the construction are found in \sect{buildingBlocks}; here we will only summarize the results.  Unlike the situation at four points, there is no single unique prefactor, but instead there are a number of prefactors that form a six-dimensional linear space. For example, for the MHV sector,  the various permutations of the following function span this space:
\be
\beta_{12345}\equiv \delta^{(8)}(Q)\frac{  \spb{1}.{2} \spb{2}.{3} \spb{3}.{4} \spb{4}.{5} \spb{5}.{1} }{4 \, \varepsilon(1,2,3,4)}\,,
\ee
where the external states are encoded in the Grassman delta function $\delta^{(8)}(Q)$, with $Q$ defined in \eqn{gdelta} using $m=5$. The denominator is the Levi-Civita invariant,  $\varepsilon(1,2,3,4)\equiv \varepsilon_{\mu\nu\rho\sigma}k_1^\mu k_2^\nu k_3^\rho k_4^\sigma={\rm Det}(k_i^{\mu})$, or the directed volume of vectors $(k_1,k_2,k_3,k_4)$.  As will be explained in detail in the next section, there is another set of equally valid  MHV prefactors given by  the various permutations of the following function:
\be
\gamma_{12}\equiv \gamma_{12345} \equiv\delta^{(8)}(Q)\frac{  \spb{1}.{2}^2 \spb{3}.{4} \spb{4}.{5} \spb{3}.{5} }{4 \, \varepsilon(1,2,3,4)}\,.
\ee
Since $\gamma_{12345}$ is totally symmetric in the three last labels, every $\gamma$ function can be uniquely specified by the two first labels.  For notational compactness, we will frequently  drop the three last labels, as done above. 
Furthermore, for higher-loop amplitudes the $\gamma_{ij}$ offers expressions  for numerators that in general are more structurally compact than those of the $\beta$, so we will use the former ones more frequently. 

Because the $\gamma$ functions satisfy the relations
\be
\sum_{i=1}^5 \gamma_{ij}=0\,, ~~~~~~ \gamma_{ij}= -\gamma_{ji}\,,
\label{gammaRel}
\ee
there are only six linearly independent $\gamma_{ij}$.   As mentioned above, the $\beta_{abcde}$ and $\gamma_{ij}$ functions are completely interchangeable,
\bea
\gamma_{12}&=&\beta_{12345}-\beta_{21345}\,, \nn \\
\beta_{12345} &=&\frac{1}{2} (\gamma_{12} + \gamma_{13} + \gamma_{14} + \gamma_{23} + \gamma_{24} + \gamma_{34}) \,.
\eea

In addition to the simple linear relations (\ref{gammaRel}), the $\gamma$'s satisfy more complicated relations when multiplied by external momentum dependent factors $s_{ij}$. For example,
\be
0=(\gamma_{12}+ \gamma_{13}) (s_{23} - s_{45}) + \gamma_{23}(s_{12} - s_{23}) + \gamma_{45}(s_{14} - s_{15})\,,
\label{twolooprel}
\ee
which through permutations of labels gives five independent linear relations. 
Such relations play an important role at higher loops as they effectively reduce the number of independent monomials that can be written down. For example, naive counting suggests that there are $6\times 5=30$ independent monomials $\gamma_{ij}s_{kl}\,$, where $s_{kl}$ are the five independent external momentum invariants. But using \eqn{twolooprel}, reduces this number to 25. Similarly, for monomials  $\gamma_{ij}s_{kl}s_{mn}$ there are $6\times 15=90$ terms, but after taking into account various  linear relations  only 66 linearly independent such terms remain~\cite{threeloopfivept}.

With the $\gamma$ (or $\beta$) universal prefactors we can write down simple ans\"{a}tze for the various master graph numerators at $L$ loops in the MHV sector,
\bea
N_i&=&\sum_{j,k,n} a_{i;jk;n}\, \gamma_{jk}\, M_n^{(L)}  \,, \nn \\
M^{(L)}&=&\left\{\prod_{l}^{L-1} m_l~ \Big|~ m_l \in \{s_{ij}, \tau_{ij} \} \right\}\,,
\label{masteransatze}
\eea
where the $a_{i;jk;n}$ are rational numbers, to be determined.  $M^{(L)}$ is the set of all independent local monomials of engineering dimension $2L-2$;  that is, the products of elementary momentum Lorentz products, denoted by $s_{ij}$ for external momenta, and $\tau_{ij}$ for internal loop momenta. At two loops, we will call the independent loop momenta in each graph $p$ and $q$, giving the possible momentum Lorentz products
\be
s_{ij}=(k_i+k_j)^2= 2 k_i \cdot k_j\,,~~~ \tau_{ip}=2 k_i \cdot p\,,~~~ \tau_{iq}=2 k_i \cdot q\,,~~~ \tau_{pq}=2 p \cdot q\,,
\ee
where $k_i$ are the external momenta (using the convention that the momenta are outgoing in any graph).  For the \MHVbar{} five-point amplitudes one can simply use the parity (or complex) conjugate expressions of  $\gamma$ (and $\beta$) functions, so we will not elaborate on this case further.

Note that the numerators in \eqn{masteransatze} are not quite local in external momenta, as the $\gamma$ (and $\beta$) functions all have a spurious divergence when the volume of  $(k_1,k_2,k_3,k_4)$ vanishes. However, the remaining factors $M_n^{(L)}$, which contains the loop momenta, are strictly local for $L>0$. Remarkably, this appears to be the perfect balance of local and nonlocal factors. As we will explicitly demonstrate in this paper and in the forthcoming~\cite{threeloopfivept},  the ans\"{a}tze~(\ref{masteransatze}) are sufficient to capture the full amplitude through at least three loops. This strongly suggests that the ans\"{a}tze sufficiently describe the five-point amplitude numerators at any loop order for the $\NeqFour$ sYM theory -- an assertion to be verified order by order in the absence of a direct proof. 
Interestingly, the ans\"{a}tze~(\ref{masteransatze}) can even be extended to tree level $L=0$ as verified in a parallel work by Br\"{o}del and one of the current authors~\cite{JJCMBroedel}.

In the final step, after having imposed the functional equations on the ans\"{a}tze, the remaining free parameters in the master numerators  are fixed by comparing against quantitative information from the theory. For example, generalized unitarity, especially maximal cuts~\cite{conformal5,compact3,JJMCHJreview}, prove to be a useful tool for this. At this point, since the duality between color and kinematics is still a conjecture, a complete verification of the constructed amplitude must of course occur; again unitarity is usually most convenient for this~\cite{GeneralizedUnitarity,GeneralizedUnitarity2}. If such verification fails then additional freedom is needed in the initial  ans\"{a}tze, and one must begin the procedure again. If some free parameters remain even after all $D$-dimensional cuts are satisfied, then these correspond to ``generalized gauge transformations''~\cite{BCJ,BCJLoop} that respect the duality and leave the full amplitude invariant. These parameters can be set to any convenient value, as they will cancel out in the full amplitude as guaranteed by the unitarity method~\cite{GeneralizedUnitarity,GeneralizedUnitarity2}.

\section{The duality-satisfying one-loop amplitudes}
\label{OneLoopSection} 

In this section, we construct the duality-satisfying one-loop five-point amplitudes of $\NeqFour$ sYM and $\NeqEight$ supergravity.  The major task in this construction is to find the proper ansatz for the numerators, that is, finding the $\beta$ and $\gamma$ functions discussed in the previous section. In this section we will assume the $\beta$ and $\gamma$ are unknown functions entering the one-loop numerators, and the effort will be to determine these.

\subsection{Diagram numerators and their ans\"{a}tze} 
\label{buildingBlocks}
\begin{figure}[t]
\begin{center}
\includegraphics[width=0.8\textwidth]{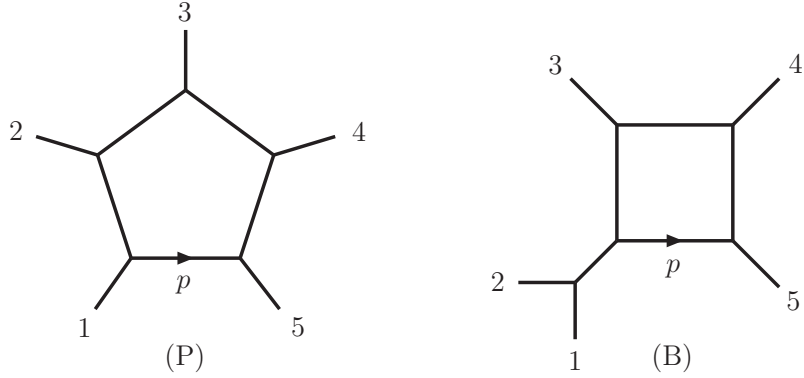}
\end{center}
\caption{\small The two diagrams that appear in the five-point one-loop amplitudes.}
\label{1L}
\end{figure}

The one-loop five-point amplitude depends on two types of cubic graphs, a pentagon and a box diagram, shown in \fig{1L}.  Their corresponding numerators will be denoted by
\be
N^{\rm (P)}(1,2,3,4,5;p)~~~~{\rm and}~~~~N^{\rm (B)}(1,2,3,4,5;p)\,, 
\ee
where the first five numeric arguments collectively encode the external state and kinematic dependence, and the last argument is the loop momentum $p$.  In addition, we could have triangle, bubble and tadpole diagrams entering the five-point one-loop amplitude; however, these are not expected to show up in maximally supersymmetric theories at one loop, so we will set the kinematic factors of these diagrams to be zero, which is later verified using unitarity cuts.

We can write down three relevant kinematic Jacobi relations for the diagrams in \fig{1L}
\bea
N^{\rm (B)}(1,2,3,4,5; p)&=&N^{\rm (P)}(1,2,3,4,5;p)-N^{\rm (P)}(2,1,3,4,5;p)\,,  \label{mastereq1}\\
0=N^{\rm (tri_2)}(1,2,3,4,5;p)&=&N^{\rm (B)}(1,2,3,4,5;p)-N^{\rm (B)}(1,2,4,3,5;p)\,, \label{mastereq2}\\
0=N^{\rm (tri_1)}(1,2,3,4,5;p)&=&N^{\rm (B)}(1,2,3,4,5;p)-N^{\rm (B)}(1,2,4,5,3;p+k_3)\,. \label{mastereq3}
\eea
As mentioned above, we immediately set the triangle numerators $N^{\rm (tri_1)}$ and  $N^{\rm (tri_2)}$ to zero, since we expect them to not be present in the amplitude.
Next, we write down the dihedral symmetry condition of the pentagon
\bea
N^{\rm (P)}(2,3,4,5,1;p+k_1)&=&N^{\rm (P)}(1,2,3,4,5;p)\,, \nn \\
N^{\rm (P)}(5,4,3,2,1;-p)&=&-N^{\rm (P)}(1,2,3,4,5;p) \,.
\label{dihedral}
\eea
Similarly we have two flip symmetries for the box
\bea
N^{\rm (B)}(2,1,3,4,5;p)&=&-N^{\rm (B)}(1,2,3,4,5;p) \,, \nn \\
N^{\rm (B)}(1,2,5,4,3;-k_1-k_2-p)&=&N^{\rm (B)}(1,2,3,4,5;p) \,.
\eea

Using \eqn{mastereq1} we can easily solve the box numerator in terms of the pentagon numerator. Thus, all the above constraints can be translated into functional equations satisfied by $N^{\rm (P)}$. To solve these we must write down an ansatz compatible with the expected structure of the amplitude. In particular, from general field-theoretic considerations, we expect there to be at least one solution for the gluonic amplitudes where the numerators are local polynomials, using polarization vectors and momentum dot products. However, here we wish to obtain a compact representation of the amplitude, thus we will avoid explicit polarization vectors as they carry a large amount of redundancy due to the mismatch of  Lorentz and little group indices. 

Having no natural building blocks for the one-loop five-point amplitude we need to be cautious with the ansatz, ensuring we parametrize all our ignorance when writing it down. Since we are looking for an amplitude where the internal loop momentum is $D$-dimensional, it makes sense to assume that the numerators are built out of local Lorentz products of the loop momentum, but the external four-dimensional states and momenta may give rise to nonlocalities. By simply counting the number of vertices in each graph we conclude that the numerators are of dimensionality 5. This implies that there can be at most five powers of loop momentum in the numerators; however, we naively expect the four-fold supersymmetry to convert four of these powers into an overall supermomentum delta function,
\be
N^{\rm (P)} \propto \delta^{(8)}(Q)\,,
\ee
leaving us with at most one power of loop momentum. 

We start with an ansatz consistent with the dihedral symmetry of the pentagon integral,
\be
N^{\rm (P)}(1,2,3,4,5;p)=\beta_{12345}+\alpha_{12345}l_1^2+\alpha_{23451}l_2^2+\alpha_{34512}l_3^2+\alpha_{45123}l_4^2+\alpha_{51234}l_5^2
\label{Pansatz}
\ee
where  $l_i=p+ k_1+k_2+\cdots k_i$ are the momenta of the five internal lines, and $\beta$ and $\alpha$ are unknown functions of the external states and momenta.   This ansatz has one more power of $p$ than we require, however, writing the amplitude in terms of  inverse propagators $l_i^2$ turns out to be more convenient for the following discussion. To make  $N^{\rm (P)}$ fully compatible with the dihedral symmetry (\ref{dihedral}) the coefficient functions must satisfy the following relations:
\be
\alpha_{abcde}=-\alpha_{baedc}\,,\hskip 1cm \beta_{abcde}=\beta_{bcdea}\,, \hskip 1cm \beta_{abcde}=-\beta_{baedc}\,,
\label{brels}
\ee
which means that there are twelve distinct $\beta$'s. 

For the box diagram we will also use an ansatz, although, in principle it is not needed since \eqn{Pansatz} implicitly generates an ansatz for $N^{\rm (B)}$. However, assuming a well-behaved ansatz for $N^{\rm (B)}$ will greatly simplify the subsequent discussion.  As is well known, the one-loop $\NeqFour$ amplitudes in $D=4$ can be represented using only scalar box integrals, where the numerators are free of loop-momentum dependence. Hence, we have reason to believe that box diagrams in the duality-satisfying $\NeqFour$ amplitudes will also be scalar integrals. Therefore, we use an ansatz free of loop momentum,
\be
N^{\rm (B)}(1,2,3,4,5;p)=\gamma_{12345}\,,
\ee
which is antisymmetric in the first two indices, and symmetric in interchange of 3rd and 5th,
\be
\gamma_{abcde}=-\gamma_{bacde}, \hskip 1cm \gamma_{abcde}=\gamma_{abedc}\,.
\label{gsym}
\ee
Next we need to impose the kinematic Jacobi relations, but first a comment about the potential redundancy of the above  ans\"{a}tze.  

The explicit appearance of loop-momentum invariants $l_i^2$ in $N^{\rm (P)}$ allows the pentagon to carry  the  exact same potential contact term as in $N^{\rm(B)}$ carried by $\gamma_{12345}$. Both such terms would correspond to a scalar box integral in the common one-loop terminology. The introduction of this apparent redundancy is, however,  well motivated.  A lesson learned at higher loops~\cite{BCJLoop}, as well as at tree level~\cite{BCJ},  is that contact terms in duality-satisfying representations have highly preferred assignment to specific cubic graphs.  In general representations, contact terms enjoy the freedom of being shuffled around, but for a duality-satisfying representation there is a delicate balance of freedom and constraints that the contact terms must obey.  So prior to constructing the one-loop duality-satisfying amplitude, one does not know whether the scalar box contributions belong to $N^{\rm (B)}$ or $N^{\rm (P)}$, or both.   In~\sect{sectionCuts}, we show that the potential redundancy is indeed a true redundancy; the duality-satisfying amplitude will allow us to make a choice that is consistent with having no scalar boxes in $N^{\rm (P)}$.

\subsection{Solving the kinematic Jacobi identities}

From  eqs.~(\ref{mastereq2}) and~(\ref{mastereq3}) we have
\be
0=\gamma_{12345}-\gamma_{12435}\,,~~~~ 0=\gamma_{12345}-\gamma_{12453}\,,
\ee
which together imply that $\gamma_{abcde}$ is symmetric in the last three indices. As no repetition of indices is allowed this means that $\gamma_{abcde}$ is completely specified by its two first indices, and therefore for notational simplicity we may simply drop the three last ones
\be
\gamma_{ab} \equiv \gamma_{abcde} \,.
\ee
From \eqn{gsym} is follows that $\gamma_{ab}=-\gamma_{ba}$, implying that there are ten distinct $\gamma_{ab}$.

Using \eqn{mastereq1} we have the following relation:
\bea
&&\gamma_{12345} =\beta_{[12]345}+\alpha_{12345}l_1^2-\alpha_{21345}(l_5^2+l_2^2-l_1^2-s_{12})+\alpha_{3[12]54}l_2^2+\alpha_{345[12]}l_3^2 \nn\\ &&
\hskip1.5cm \null+\alpha_{45[12]3}l_4^2+\alpha_{5[12]34}l_5^2 \, ,
\eea
where the square brackets, $[\,]$, signify antisymmetrization of the arguments, and where we used the relation $l_1^2 |_{k_1\leftrightarrow k_2}=(p+k_2)^2=l_5^2+l_2^2-l_1^2-s_{12}$. The $l_i^2$ are independent variables so the equation can be decomposed into components, giving the relations
\bea
&&\alpha_{(12)345}=0\,,~~~\alpha_{345[12]}=0\,,~~~\alpha_{45[12]3}=0\,, \label{asym1} \\
&& \alpha_{5[12]34}-\alpha_{21345}=0\,,~~~ \alpha_{3[12]54}-\alpha_{21345}=0 \,, \label{asym2}  \\
&& \gamma_{12345}=\beta_{[12]345}+s_{12}\alpha_{21345}\,, \label{gsol} 
\eea
where the round brackets, $(\,)$, mean symmetrization of the indices.
The first row (\ref{asym1}) implies that $\alpha_{abcde}$ is antisymmetric in the two first indices, and symmetric in the last three ones, so just like for $\gamma$ we can drop the last three arguments $\alpha_{ab} \equiv \alpha_{abcde}$. Using this notation the second row (\ref{asym2}) can be summarized as follows:
\be
\alpha_{ab}=\alpha_{a1}+\alpha_{1b} \,,
\ee
implying that only four $\alpha$'s are independent, namely $\alpha_{12},\alpha_{13},\alpha_{14},\alpha_{15}$.

The last equation (\ref{gsol}) solves all the $\gamma$'s in terms of the $\beta$'s and $\alpha$'s, but it also implies that $\beta_{[ab]cde}$
is totally symmetric in the last three indices, just like $\alpha$ and $\gamma$. This means that not all twelve $\beta$ are independent, since {\it e.g.} $\beta_{[ab][cd]e}=0$. This double-commutivity constraint taken with \eqn{brels} can be recast as
\be
\beta_{[12]}+\beta_{[13]}+\beta_{[14]}+\beta_{[15]}=0\,,
\label{betaid}
\ee
where we again, without loss of information, dropped the last three indices $\beta_{[ab]} \equiv \beta_{[ab]cde}=\beta_{abcde}-\beta_{bacde}$. We may solve this, and similar equations obtained by permutations, by eliminating all variables $\beta_{[i5]}$. This results in six independent variables: $\beta_{[12]},\beta_{[13]},\beta_{[14]},\beta_{[23]},\beta_{[24]}$ and $\beta_{[34]}$. Indeed, we can express $\beta_{12345}$ in terms of these:
\be
\beta_{12345}=\frac{1}{2}(\beta_{[12]}+\beta_{[13]}+\beta_{[14]}+\beta_{[23]}+\beta_{[24]}+\beta_{[34]})\,,
\ee
as can be shown by combining \eqn{brels} with $\beta_{[ab][cd]e}=0$.

Thus, we conclude that after solving all the Jacobi relations, the remaining unconstrained degrees of freedom are ten in total:  four distinct permutations of the $\alpha$ parameters and six permutations of the $\beta$ parameters. In the next section, we will find explicit expressions for these.

\subsection{Fixing the remaining parameters from unitarity cuts}
\label{sectionCuts}

To get the final expressions for $\beta$ and $\alpha$  we match to a quadruple cut~\cite{BCF} of the pentagon and box. That is, the cut where all $l_i^2=0$ except $l_1^2\neq0$ are on shell. We find the following expression for the cut applied to the current amplitude ansatz:
\be
\frac{\beta_{12345}+\alpha_{12}l_1^2}{l_1^2}+ \frac{\gamma_{12}}{s_{12}}= \frac{\beta_{12345}}{(p+k_1)^2}+ \frac{\beta_{[12]345}}{s_{12}} \,.
\label{quadcut}
\ee
Interestingly, after using \eqn{gsol} to obtain the right-hand side, the $\alpha_{12}$ parameter completely cancels out between the two diagrams. Indeed, this cut shows that it cancels out in the full amplitude.  The $\alpha_{12}$ contribution to the amplitude is always in the form of a scalar box diagram, {\it i.e.} the inverse propagator $l_1^2$ cancels one of the pentagon edges, giving a box. Since the quadruple cut does not detect this box, this means that $\alpha_{12}$ does not contribute to the amplitude. Thus, a solution consistent with the unitarity cuts is
\be
\alpha_{ab}=0.
\label{acondit}
\ee
It should be stressed that this is a choice, so there is the possibility of making other nonzero choices resulting in alternative one-loop duality-satisfying representations -- but this choice is clearly the simplest one as it removes the loop-momentum dependence in the pentagon numerator. This choice is also consistent with the generalizations of duality-satisfying representations at higher loops, as we will see.

After specifying the cut to $D=4$  \eqn{quadcut} gives us two equations, as there are two solutions to $p$ in this dimension~\cite{BCF}. For the MHV configuration we have the following loop-momentum solution:
\bea
p&=&\frac{(k_1+k_2)|3\rangle\langle 5|}{\spa{3}.{5}}\,, \\
(p+k_1)^2&=&\frac{\langle 5| k_1(k_1+k_2)|3\rangle}{\spa{3}.{5}}=\frac{\spa{5}.{1}\spb{1}.{2}\spa{2}.{3}}{\spa{3}.{5}}\,.
\eea
The \MHVbar{} configuration is trivially obtained by parity conjugation $\langle \rangle \leftrightarrow []$ of the above expressions. The two equations we need to solve for the MHV amplitude are
\bea
i\delta^{(8)}(Q)\frac{s_{34}s_{45}}{\spa{1}.{2}\spa{2}.{3}\spa{3}.{4}\spa{4}.{5}\spa{5}.{1}}&=& \beta_{12345}\frac{\spa{3}.{5}}{\spa{5}.{1}\spb{1}.{2}\spa{2}.{3}}+\frac{\beta_{[12]345}}{s_{12}} \label{MHVsol}\,, \\
0&=& \beta_{12345}\frac{\spb{3}.{5}}{\spb{5}.{1}\spa{1}.{2}\spb{2}.{3}}+ \frac{\beta_{[12]345}}{s_{12}}\,, \label{MHVbarsol}
\eea
where the first expression is twice the value of the well-known box coefficient~\cite{MHVoneloop}, and the second equation states that the quadruple cut vanishes on the complex conjugate solution. This happens because the cut effectively involves on-shell three-point vertices, which only has support on one chiral branch of the five-point amplitude~\cite{BCF}. The reason we use twice the value of the box coefficient in the first equation is that the box coefficient is usually computed as the average of the two above solutions~\cite{BCF}.

Solving for $\beta_{12345}$ is now straightforward.  We take the difference of the two equations (\ref{MHVsol}) and (\ref{MHVbarsol}), giving
\be
\beta_{12345}=i \delta^{(8)}(Q)\frac{\spb{1}.{2}\spb{2}.{3}\spb{3}.{4}\spb{4}.{5}\spb{5}.{1}}{\spa{1}.{2}\spb{2}.{3}\spa{3}.{5}\spb{5}.{1}-\spb{1}.{2}\spa{2}.{3}\spb{3}.{5}\spa{5}.{1}}=\delta^{(8)}(Q)\frac{\spb{1}.{2}\spb{2}.{3}\spb{3}.{4}\spb{4}.{5}\spb{5}.{1}}{4\, \varepsilon(1,2,3,4)}\,,
\label{betaform}
\ee
 where $\varepsilon(1,2,3,4)={\rm Det}(k_i^{\mu})$ is the Levi-Civita invariant, or the directed volume of vectors $(k_1,k_2,k_3,k_4)$.
 
Using this expression for $\beta_{12345}$ it is trivial to check that also the second equation (\ref{MHVbarsol}) is satisfied. This concludes the four-dimensional unitarity checks of the obtained amplitude, since for one-loop $\NeqFour$ amplitudes the quadruple cuts completely specify the amplitude~\cite{MHVoneloop,BCF}. 

Now we can substitute the solution into the box numerator,
\bea
N^{\rm (B)}=\gamma_{12}&=&\beta_{12345}-\beta_{21345}=\delta^{(8)}(Q)\frac{\spb{1}.{2}^2\spb{3}.{4}\spb{4}.{5}\spb{3}.{5}}{4\, \varepsilon(1,2,3,4)}\,,
\label{gammaform}
\eea
where we have used a Schouten identity $\spb{5}.{1}\spb{2}.{3}-\spb{5}.{2}\spb{1}.{3}=\spb{1}.{2}\spb{3}.{5}$. Indeed, $\gamma_{12}=\gamma_{12345}$ is antisymmetric in the first two indices, and totally symmetric in the last three ones, as demanded by the duality.  We can also check that $\beta_{12345}$ satisfies the duality constraints: except for the $\varepsilon(1,2,3,4)$ factor in the denominator it has obvious dihedral symmetry. The fact that there are only four independent external momenta means that $\varepsilon(1,2,3,4)=\varepsilon(2,3,4,5)$ and thus the denominator also respects the dihedral symmetry.

Since $\alpha_{ij}=0$, it now follows from \eqn{gsol} that $\gamma_{ij}=\beta_{[ij]}$, and we should have the following relations:
\be
\beta_{12345}=\frac{1}{2}(\gamma_{12}+\gamma_{13}+ \gamma_{14}+ \gamma_{23}+ \gamma_{24}+ \gamma_{34}) \,,
\ee
and
\be 
\sum_{i=1}^5 \gamma_{ij}=0\,,
\ee
which is a rewrite of \eqn{betaid} using $\gamma_{ij}=-\gamma_{ji}$. Indeed, these relations completely agree with the explicit forms in (\ref{betaform}) and (\ref{gammaform}). This shows that the obtained amplitude solution obeys the duality between color and kinematics.  Note that the linear relations and properties satisfied by $\beta$ and $\gamma$  follow from the duality; indeed, except for the choice of setting $\alpha_{ij}=0$, we arrived at these relations even before having imposed the unitarity cuts.

Finally, we note that the resulting duality-satisfying one-loop five-point $\NeqFour$ sYM amplitude is given in terms of only scalar integrals; remarkably, no loop momentum is needed in the pentagon numerator. Given this simple form for the amplitude it is no surprise that this representation has been found before. In ref.~\cite{CachazoLeadingSingularityAndCalcs}, Cachazo obtains a very similar integral form for this amplitude using only a scalar pentagon and a scalar box. In an even older paper, Bern and Morgan~\cite{BernMorgan} gives the same amplitude implicitly in the disguise of a one-loop all-plus Yang-Mills amplitude, which is simply related to the $\NeqFour$ sYM through a dimension-shifting formula~\cite{DimShift}. We have checked that the $N^{\rm (P)}$ and $N^{\rm (B)}$ numerators agree with the prefactors of the integrals given in these papers, showing that the representations are the same diagram by diagram.

\subsection{The one-loop five-point MHV amplitudes}

Here we give the complete one-loop five-point MHV amplitudes of $\NeqFour$ sYM and $\NeqEight$ supergravity. The external momenta and states are defined in $D=4$ and the internal loop integration is for any dimension where the maximally supersymmetric theories are defined. The $\NeqFour$ sYM amplitude is
\be
{\cal A}_5^{(1)} =  i g^{5} \, \sum_{S_5} \, 
\Bigl( 
{\frac{1}{10}}\beta_{12345}C^{(\rm P)} I^{(\rm P)}+{\frac{1}{4}}\gamma_{12} C^{(\rm B)}I^{(\rm B)}
\Bigr) \,, 
\label{OneLoopSYMAmplitude}
\ee
where $g$ is the coupling constant, and the sum is over all 120 permutations, $S_5$, of the external leg labels; the symmetry factors 1/10 and 1/4 compensate for the overcount in this sum. Functions $\beta_{12345}$ and $\gamma_{12}$ are given in (\ref{betaform}) and (\ref{gammaform}). 
The integrals are given by
\bea
I^{(\rm P)}&=&\int \frac{d^Dp}{(2\pi)^D} \frac{1}{p^2(p+k_1)^2(p+k_1+k_2)^2(p-k_4-k_5)^2(p-k_5)^2}\,, \nn \\
I^{(\rm B)}&=&\frac{1}{s_{12}}\int \frac{d^Dp}{(2\pi)^D} \frac{1}{p^2(p+k_1+k_2)^2(p-k_4-k_5)^2(p-k_5)^2}\,,
\label{integrals}
\eea
and the color factors are
\bea
C^{(\rm P)}&=& \f^{g a_1b}\f^{b a_2 c}\f^{c a_3 d}\f^{d a_4 e }\f^{e a_5 g} \,, \nn \\
C^{(\rm B)}&=& \f^{a_1a_2 b}\f^{b c g}\f^{c a_3 d}\f^{d a_4 e }\f^{e a_5 g} \,,
\eea
where $a_i$ are the external color labels.

The $\NeqEight$ supergravity amplitude is given by
\be
{\cal M}_5^{(1)} = - \left(\frac{\kappa}{2}\right)^{5} \, \sum_{S_5} \, 
\Bigl(
{\frac{1}{10}}\beta_{12345} {\tilde\beta}_{12345}\, I^{(\rm P)}+{\frac{1}{4}}\gamma_{12} {\tilde\gamma}_{12} I^{(\rm B)}
\Bigr) \,, 
\label{OneLoopSGAmplitude}
\ee
where $\kappa$ is the gravity coupling constant. As above, the sum is over all 120 permutations, $S_5$, of the external leg labels, and the integrals are given above (\ref{integrals}). The ${\tilde\beta}_{12345}$ and ${\tilde\gamma}_{12}$ are the same as the untilded functions, (\ref{betaform}) and (\ref{gammaform}),  except that the $SU(4)$ $R$-symmetry indices are shifted $\eta_i^A \rightarrow \eta_i^{A+4}$ in the super momentum delta function $\delta^{(8)}(Q)$, since they are embedded in the ``second half'' of $SU(8)$.

The above one-loop amplitudes have been verified using unitarity cuts in both general dimension $D$ and in $D=4$. The four-dimensional contributions are detected by a quadruple cut~\cite{MHVoneloop,BCF}, and in addition we have performed the $D$-dimensional pentacuts applied to the pentagon (P) and to the box (B) (including cutting the external propagator $s_{12}=0$)  . The  one-loop five-point $\NeqFour$ sYM amplitude also matches the known expressions in the literature~\cite{CachazoLeadingSingularityAndCalcs,BernMorgan}.

\subsection{UV divergences at one loop}

Using the one-loop amplitudes in \eqn{OneLoopSYMAmplitude} and \eqn{OneLoopSGAmplitude} we can easily compute the logarithmic ultraviolet divergences that first occur in $D=8$ for both theories. For dimensions $D<8$ both theories are expected to be finite at one loop~\cite{GSB}, as is manifest for the amplitudes (\ref{OneLoopSYMAmplitude}) and (\ref{OneLoopSGAmplitude}). The $D=8-2\eps$ divergence arises from the box diagram only, with the known result~\cite{GSB,Neq44np}
\be
I^{(\rm B)}\Bigl|_{\rm UV~pole}=\frac{i}{6(4\pi)^4\epsilon}\frac{1}{s_{12}}\,.
\ee
Using this, and letting the subleading $I^{(\rm P)}$ vanish in the critical dimension, we get for $SU(N_c)$ $\NeqFour$ sYM the divergence
\bea
{\cal A}_5^{(1)}\Bigr|_{\rm UV}
&=& -g^{5} \,  \frac{1}{6(4\pi)^4\epsilon}\Bigl[
N_c \Tr_{12345}\Bigl(\frac{\gamma_{12}}{s_{12}}+\frac{\gamma_{23}}{s_{23}}+\frac{\gamma_{34}}{s_{34}}+\frac{\gamma_{45}}{s_{45}}+\frac{\gamma_{51}}{s_{15}}\Bigr) \nn \\ &&
\null~~~~~~~~~~~~~~~+6\Tr_{123}\Tr_{45}\Bigl(\frac{\gamma_{12}}{s_{12}}+\frac{\gamma_{23}}{s_{23}}+\frac{\gamma_{31}}{s_{13}}\Bigr)+{\rm~perms}
\Bigr]\,,
\label{OneLoopSYMDiv}
\eea
where $\Tr_{12\cdots n}=\Tr(T^{a_1}T^{a_2} \cdots T^{a_n})$ encodes the gauge-group trace structures. The remaining trace structures, hidden in the ``$+$ perms,'' can be obtained from the given ones by using crossing symmetry.

For the $\NeqEight$ supergravity divergence we have
\be
{\cal M}_5^{(1)}\Bigr|_{\rm UV} \hskip-1mm =
-i \left(\frac{\kappa}{2}\right)^{5} \,  \frac{1}{6(4\pi)^4\epsilon}\Bigl[
\frac{\gamma_{12}^2}{s_{12}}+\frac{\gamma_{13}^2}{s_{13}}+\frac{\gamma_{14}^2}{s_{14}}+\frac{\gamma_{15}^2}{s_{15}}+\frac{\gamma_{23}^2}{s_{23}}+\frac{\gamma_{24}^2}{s_{24}}+\frac{\gamma_{25}^2}{s_{25}}+\frac{\gamma_{34}^2}{s_{34}}+\frac{\gamma_{35}^2}{s_{35}}+\frac{\gamma_{45}^2}{s_{45}}
\Bigr]\,, 
\label{OneLoopSGDiv}
\ee
where we for convenience defined $\gamma_{ij}^2\equiv \gamma_{ij} {\tilde \gamma}_{ij}=  \gamma_{ij} (\gamma_{ij} |_{\eta_i^A \rightarrow \eta_i^{A+4}})$. 

The forms of these five-point one-loop divergences are compatible with the logarithmic divergences observed for the four-point  amplitudes of the two theories~\cite{GSB,Neq44np}. Indeed, one can easily recover the corresponding four-point divergences in any of the factorization channels $s_{ij}\rightarrow0$. Seeing no additional local structure at five points we expect that the counterterms to these divergences should be the same as those at the four-point level, namely, of the schematic forms $F^4$ and $R^4$ for gluon and graviton components, respectively.

\section{Two-loop five-point solution}
\label{twoloopsection}

\begin{figure}[t]
\begin{center}
\includegraphics[width=0.85 \textwidth]{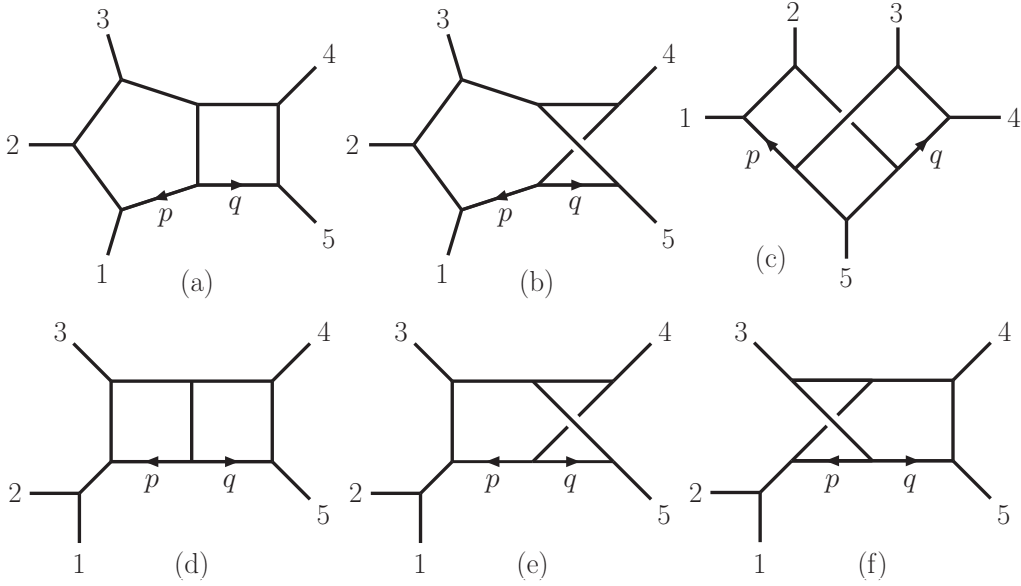}
\end{center}
\caption{\label{2L1}\small The six diagrams that appear in the five-point two-loop amplitudes.}
\end{figure}

Now we want to analyze the five-point two-loop amplitude. We will assume that the diagram basis for $\NeqFour$ sYM involves the six graphs in \fig{2L1}. These diagrams are obtained by eliminating all triangles, bubbles and tadpoles from a generic $D$-dimensional basis, and then additionally eliminating any diagrams with a three-point two-loop subgraph. We will use the canonical notation
\be
N^{(x)}=N^{(x)}(1,2,3,4,5; p,q)
\ee
to denote the numerators of the six diagrams in \fig{2L1}, where the first five arguments encode both external states and external momenta, and $p$ and $q$ are the two loop momenta.

%
%

To simplify the analysis we will note the following property: in an amplitude representation free of triangle subgraphs, the diagram numerators have to be totally symmetric with respect to permutations of any four legs that connect to a box subgraph. This follows from the kinematic Jacobi relations, since, triangles are obtained from the antisymmetrization of any two legs in a box diagram. The absence of triangles is then equivalent to requiring total symmetry of the box numerators.  This explains why the numerator of diagram (B) in \fig{1L} is totally symmetric in legs 3, 4 and 5. 
And, for two multiloop diagrams, which only differ by the ordering of legs of a box subgraph, it follows that they have the same numerator. At two loops this property implies the following constraints on the numerators:
\be
N^{(\rm a)}=N^{(\rm b)}, \hskip 2cm N^{(\rm d)}=N^{(\rm e)}=N^{(\rm f)}\,.
\label{trivialrel}
\ee
This can easily be seen in \fig{2L1}: diagram (a) and (b) only differ by the edge connections of the rightmost one-loop subgraph, which is a box. Similarly, (d) differs from (e) by connections in the rightmost one-loop subgraph, and (d) differs from (f) by  connections in the leftmost one-loop subgraph, both are boxes.

Further, the remaining undetermined numerators $N^{\rm (a)}$, $N^{\rm (c)}$ and $N^{\rm (d)}$ are interlocked by the two kinematic Jacobi relations,
\bea
N^{\rm (c)}(1,2,3,4,5; p,q)&=&N^{\rm (a)}(1,2,5,4,3; p,k_{3,4} -q)-N^{\rm (a)}(5,4,3,1,2; k_5+q,k_{1,2}-p)\,, \nn \\
N^{\rm (d)}(1,2,3,4,5; p,q)&=&N^{\rm (a)}(1,2,3,4,5; p,q)-N^{\rm (a)}(2,1,3,4,5; p,q)\,,
\label{bcjsol}
\eea
where $k_{i,j}=k_i+k_j$. There are many more kinematic Jacobi relations that one can write down but these two are sufficient for reducing the system to only one unknown numerator. It is clear that $N^{\rm (a)}$ determines the numerators of all other diagrams, thus, all we need to do is to find the explicit expression for this master numerator.  Alternatively, we could have used diagram (c) as the master diagram, as the following Jacobi relation entails:
\be
N^{(\rm b)}(1,2,3,4,5; p,q)=-N^{(\rm c)}(1,2,5,3,4; p,k_{3,5}-q)-N^{(\rm c)}(1,2,4,3,5; p,k_{3,4}+p+q)\,.
\label{alternativeJacobi}
\ee
This numerator will have a more complicated function structure, as graph (c) contains no box subdiagrams. Therefore, it is strategically wiser to choose the planar graph (a) as the master.

We may also study automorphism symmetries of the diagrams; from \fig{2L1} we see that the numerators should satisfy the following self-relations:
\bea
N^{\rm (a)}(1,2,3,4,5;p,q)&=&-N^{\rm (a)}(3,2,1,5,4; k_{1,2,3}-p,k_{4,5}-q)\,,\nn \\
N^{\rm (b)}(1,2,3,4,5;p,q)&=&-N^{\rm (b)}(3,2,1,4,5; k_{1,2,3}-p,k_5-q)\,,\nn \\
N^{\rm (b)}(1,2,3,4,5;p,q)&=&N^{\rm (b)}(1,2,3,5,4; p,p+q+k_4)\,,\nn \\
N^{\rm (c)}(1,2,3,4,5;p,q)&=&-N^{\rm (c)}(4,3,2,1,5; q,p)\,,\nn \\
N^{\rm (c)}(1,2,3,4,5;p,q)&=&N^{\rm (c)}(3,4,1,2,5; k_{3,4}-q,k_{1,2}-p)\,,\nn \\
N^{\rm (d)}(1,2,3,4,5;p,q)&=&-N^{\rm (d)}(2,1,3,4,5; p,q)\,,\nn \\
N^{\rm (e)}(1,2,3,4,5;p,q)&=&-N^{\rm (e)}(2,1,3,4,5; p,q)\,,\nn \\
N^{\rm (e)}(1,2,3,4,5;p,q)&=&N^{\rm (e)}(1,2,3,5,4; p,p+q+k_4)\,,\nn \\
N^{\rm (f)}(1,2,3,4,5;p,q)&=&-N^{\rm (f)}(2,1,3,4,5; p,q)\,,\nn \\
N^{\rm (f)}(1,2,3,4,5;p,q)&=&-N^{\rm (f)}(1,2,3,5,4; k_{1,2}-p,k_{4,5}-q)\,.
\label{autosym}
\eea
These are in fact all the independent automorphism symmetries of each diagram. 

After having written down some of the needed functional equations, we proceed by finding a suitable solution. The calculation will actually be simpler than at one loop since a well-behaved ansatz for the master numerator is readily available by recycling the one-loop results. As discussed in \sect{GeneralStructureSection} we expect the full state dependence of the two-loop amplitude to be captured by the following six $\gamma$ parameters,
\be
\gamma_{12},~~~ \gamma_{13},~~~\gamma_{14},~~~ \gamma_{23},~~~ \gamma_{24},~~~ \gamma_{34}\,.
\ee
The ansatz is then
\be
N^{\rm (a)}= \gamma_{12} \, m_1+\gamma_{13}\,m_2+\gamma_{14}\,m_3+\gamma_{23}\,m_4+ \gamma_{24}\,m_5+ \gamma_{34}\,m_6\,,
\ee
where the $m_j$ are local state-independent objects. By dimensional analysis they are quadratic in momenta; thus, we may parametrize them as
\be
m_j=a_{1j} s_{12}+a_{2j} s_{13}+a_{3j} s_{14}+a_{4j} s_{23}+a_{5j} s_{24}+a_{6j}  \tau_{1p}+a_{7j}  \tau_{2p}+a_{8j}  \tau_{3p}+a_{9j} \tau_{4p}\,,
\ee
where we have assumed that the numerator does not depend on the momenta of the box subdiagram, and is at most linear in the momentum of the pentagon subdiagram. The parameters $a_{ij}$ are constant rational numbers, accounting for in total $6\times9= 54$ undetermined parameters. However, because of the relations (\ref{twolooprel}), there is a slight over-parametrization of the function space. We may consequently set five parameters to zero, {\it e.g.}
\be
a_{45}=  a_{26}= a_{36}= a_{46}=a_{56}= 0\,.
\ee 
Now we have  $49$ undetermined parameters.

First we enforce the kinematic Jacobi relation in \eqn{alternativeJacobi} using the solution \eqn{bcjsol} for $N^{(c)}$. This relation contains 44 independent constraints, reducing the ansatz down to only five free parameters. 
We can fix one additional parameter by requiring that  $N^{\rm (a)}$ has the correct automorphism symmetry, given by the first line of \eqn{autosym}. Now we have only four free parameters. The remaining equations in (\ref{autosym}) are automatically satisfied by this four-parameter ansatz, and remarkably, all possible kinematic Jacobi relations that one can write down are satisfied. Thus no more parameters can be fixed without inputting some quantitative information, for example, from a unitarity cut.  

%
\begin{figure}[t]
\begin{center}
\includegraphics[width=0.85 \textwidth]{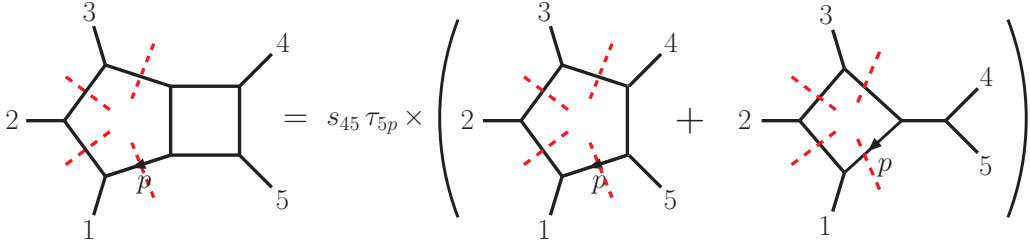}
\end{center}
\caption{\label{BoxCutFivePt}\small A boxcut for the two-loop pentabox graph excises the box (dropping the propagators) and writes the result in terms of two one-loop diagrams times a kinematic factor.}
\end{figure}
%

We will do a ``boxcut''~\cite{Neq44np} on diagram (a) to fix the remaining four parameters. That is, we will excise the one-loop box diagram in (a) using the on-shell conditions $p^2=(p-k_1)^2=(p-k_1+k_2)^2=(p-k_1-k_2-k_3)^2=0$, and then map the cut to a linear combination of one-loop numerators, see \fig{BoxCutFivePt}. The resulting expression for the cut is 
\be
N^{(\rm a)}\Big|_{\rm cut}= s_{45}\, \tau_{5p}  \left(\frac{\beta_{12345}}{\tau_{5p}}+ \frac{\gamma_{45}}{s_{45}} \right) \,,
\ee
where the kinematic rules  $\{\tau_{1p}\rightarrow0,  \tau_{2p} \rightarrow s_{12}, \tau_{3p} \rightarrow s_{45}-s_{12} ,\tau_{4p} \rightarrow -\tau_{5p} - s_{45} \}$ should be imposed on this unitarity cut.
This cut equation fixes the remaining four parameters. The solution, in terms of the 49 ansatz parameters, is given by
\bea
&&
a_{12} = a_{21} = a_{41} = \Frac{1}{2}\,,~ a_{14} = \Frac{3}{4}\,, ~ a_{93} = a_{95} = a_{96} = -1\,, ~ a_{61} = a_{62} = a_{74} = -\Frac{1}{4}, \nn \\ && 
a_{11} = a_{22} = a_{24}= a_{42} = a_{44} = a_{71} = a_{82} = a_{84} = \Frac{1}{4}\,,  \nn \\&&
a_{63} = a_{65} = a_{66} = a_{73} = a_{75} = a_{76} = a_{83} = a_{85} = a_{86} = -\Frac{1}{2}\,, \nn \\&&
a_{13} = a_{15} = a_{16} = a_{23} = a_{25} = a_{31} = a_{32} = a_{33} = a_{34} = a_{35} = a_{43} = a_{51} = a_{52} = a_{53} \nn \\&&
~~~~ = a_{54} = a_{55} = a_{64} = a_{72} = a_{81} = a_{91} = a_{92} = a_{94} = 0\,.
\eea
After some cleanup, using momentum identities and the relations (\ref{twolooprel}) and (\ref{gammaRel}), the numerator of diagram (a)  is given by
\bea
N^{(\rm a)}(1,2,3,4,5; p,q)&=&\frac{1}{4} \Bigl(\gamma_{12} (2 s_{45}- s_{12} + \tau_{2p} - \tau_{1p})+ \gamma_{23} (s_{45}  + 2 s_{12} - \tau_{2p} + \tau_{3p})  \nn \\&& 
~~~~\null + 2 \gamma_{45} (\tau_{5p} - \tau_{4p})+ \gamma_{13} (s_{12}  + s_{45} - \tau_{1p} + \tau_{3p}) \Bigr)\,.
\eea
The other five graph numerators can easily be obtained through (\ref{bcjsol}) and (\ref{trivialrel}). For convenience they are also given in \tab{NumeratorTable}.

\begin{table*}\caption{The numerator factors of the integrals in
\fig{2L1}. The first column labels the integral, the second
column the numerator factor for $\NeqFour$ super-Yang-Mills
theory. The squares of these, or more accurately their double copies, are the numerator factors for
$\NeqEight$ supergravity. 
\label{NumeratorTable} }
\vskip .4 cm
\begin{center}
\begin{tabular}{||c|c||}
\hline
 ${\cal I}^{(x)}$ & $\NeqFour$ Super-Yang-Mills ($\sqrt{\NeqEight~{\rm supergravity}}$) numerator \\
\hline
\hline
(a),(b) & 
$\frac{1}{4} \Bigl(\gamma_{12} (2 s_{45}- s_{12} + \tau_{2p} - \tau_{1p})+ \gamma_{23} (s_{45}  + 2 s_{12} - \tau_{2p} + \tau_{3p})   $\\
&$\null 
\hskip 2.3cm + 2 \gamma_{45} (\tau_{5p} - \tau_{4p})+ \gamma_{13} (s_{12}  + s_{45} - \tau_{1p} + \tau_{3p}) \Bigr)$
\\
\hline 
(c) &
$\frac{1}{4} \Bigl( \gamma_{15} (\tau_{5p} - \tau_{1p})+ \gamma_{25} (s_{12} - \tau_{2p} + \tau_{5p}) + \gamma_{12} (s_{34} + \tau_{2p} - \tau_{1p} + 2 s_{15} + 2 \tau_{1q} -  2 \tau_{2q}) $\\
&$\null 
~~+ \gamma_{45}  (\tau_{4q} - \tau_{5q}) - \gamma_{35} (s_{34} - \tau_{3q} + \tau_{5q}) + \gamma_{34} (s_{12} + \tau_{3q} - \tau_{4q} + 2 s_{45} + 2 \tau_{4p} - 2 \tau_{3p} ) \Bigr)$ 
\\
\hline
(d)-(f) & $  \gamma_{12} s_{45}-\frac{1}{4} \Bigl(2 \gamma_{12} + \gamma_{13} - \gamma_{23}\Bigr) s_{12} $
 \\
\hline
\end{tabular}
\end{center}
\end{table*}

\subsection{The two-loop five-point MHV amplitudes}

Here we give the complete two-loop five-point MHV amplitudes of $\NeqFour$ sYM and $\NeqEight$ supergravity. The external momenta and states are defined in $D=4$ and the internal loop integration is for any dimension where the maximally supersymmetric theories are defined. The $\NeqFour$ sYM amplitude is
\be
{\cal A}_5^{(2)} =  - g^{7} \, \sum_{S_5} \, 
\Bigl(
{\frac{1}{2}}\I^{(\rm a)}+{\frac{1}{4}}\I^{(\rm b)}+{\frac{1}{4}}\I^{(\rm c)}
+{\frac{1}{2}}\I^{(\rm d)}+{\frac{1}{4}}\I^{(\rm e)}+{\frac{1}{4}}\I^{(\rm f)}
\Bigr) \,, 
\label{TwoLoopSYMAmplitude}
\ee
where $g$ is the coupling constant, and the sum is over all 120 permutations, $S_5$, of the external leg labels; the symmetry factors 1/2 and 1/4 compensate for the overcount in this sum.
The integrals are given by
\be
{\cal I}^{(x)}=\int \frac{d^Dp}{(2\pi)^D} \frac{d^Dq}{(2\pi)^D} \frac{C^{(x)}N^{(x)}(1,2,3,4,5;p,q)}{l_1^2\,l_2^2\,l_3^2\,l_4^2\,l_5^2\,l_6^2\,l_7^2\,l_8^2}\,, 
\label{integrals2}
\ee
where the $l_i$ are linear combinations of $k_i,p$ and $q$, according to the graph structure of each diagram in \fig{2L1} [for diagrams (d), (e) and (f) one of the $1/l_i^2$ is an external propagator $1/s_{12}$].  The numerators $N^{(x)}$ are given in \tab{NumeratorTable}. 
The color factors are
\bea
C^{(\rm a)}&=&c_{(4, 10, 8)} c_{(5, 7, 10)} c_{(6, 1, 12)} c_{(7, 6, 9)} c_{(8, 9, 11)} c_{(11, 13, 3)} c_{(12, 2, 13)}\,,\nn \\ 
C^{(\rm b)}&=& c_{(4, 9, 10)} c_{(5, 7, 8)} c_{(6, 1, 12)} c_{(8, 9, 11)} c_{(10, 7, 6)} c_{(11, 13, 3)} c_{(12, 2, 13)}\,,\nn \\ 
C^{(\rm c)}&=&c_{(1, 6, 8)} c_{(2, 12, 8)} c_{(6, 9, 11)} c_{(7, 4, 13)} c_{(10, 9, 5)} c_{(11, 13, 3)} c_{(12, 7, 10)}\,,\nn \\
C^{(\rm d)}&=&c_{(4, 10, 8)} c_{(5, 7, 10)} c_{(6, 13, 12)} c_{(7, 6, 9)} c_{(8, 9, 11)} c_{(11, 13, 3)} c_{(12, 2, 1)}\,,\nn \\
C^{(\rm e)}&=&c_{(4, 10, 8)} c_{(5, 9, 7)} c_{(6, 13, 12)} c_{(7, 10, 6)} c_{(8, 9, 11)} c_{(11, 13, 3)} c_{(12, 2, 1)}\,,\nn \\
C^{(\rm f)}&=&c_{(2, 1, 8)} c_{(6, 9, 7)} c_{(7, 13, 5)} c_{(8, 6, 11)} c_{(10, 3, 9)} c_{(11, 12, 10)} c_{(12, 4, 13)}\,,
\eea
where we use the notation $c_{(i,j,k)}\equiv \f^{a_i a_j a_k}$ for the structure constants, and $a_{i\le5}$ are the external color labels.

The $\NeqEight$ supergravity amplitude is given by
\be
{\cal M}_5^{(2)} = 
-i \left(\frac{\kappa}{2}\right)^{7} \, \sum_{S_5} \, 
\Bigl(
{\frac{1}{2}}I^{(\rm a)}+{\frac{1}{4}}I^{(\rm b)}+{\frac{1}{4}}I^{(\rm c)}
+{\frac{1}{2}}I^{(\rm d)}+{\frac{1}{4}}I^{(\rm e)}+{\frac{1}{4}}I^{(\rm f)}
\Bigr) \,, 
\label{TwoLoopSGAmplitude}
\ee
where $\kappa$ is the gravity coupling constant, and as above the sum is over all 120 permutations, $S_5$, of the external leg labels. The integrals are given by
\be
I^{(x)}=\int \frac{d^Dp}{(2\pi)^D} \frac{d^Dq}{(2\pi)^D} \frac{N^{(x)}(1,2,3,4,5;p,q) \tilde{N}^{(x)}(1,2,3,4,5;p,q)}{l_1^2\,l_2^2\,l_3^2\,l_4^2\,l_5^2\,l_6^2\,l_7^2\,l_8^2}\,, 
\ee
where second numerator copy has shifted $R$-symmetry indices $\tilde{N}^{(x)}(1,2,3,4,5;p,q)=N^{(x)}(1,2,3,4,5;p,q)|_{\eta_i^A \rightarrow \eta_i^{A+4}}\, $, and  $N^{(x)}$ are given in \tab{NumeratorTable}. As above the $l_i$'s dependence on $k_i,p$ and $q$ follows from each diagram in \fig{2L1}.

\begin{figure}[t]
\centerline{\epsfxsize 6 truein \epsfbox{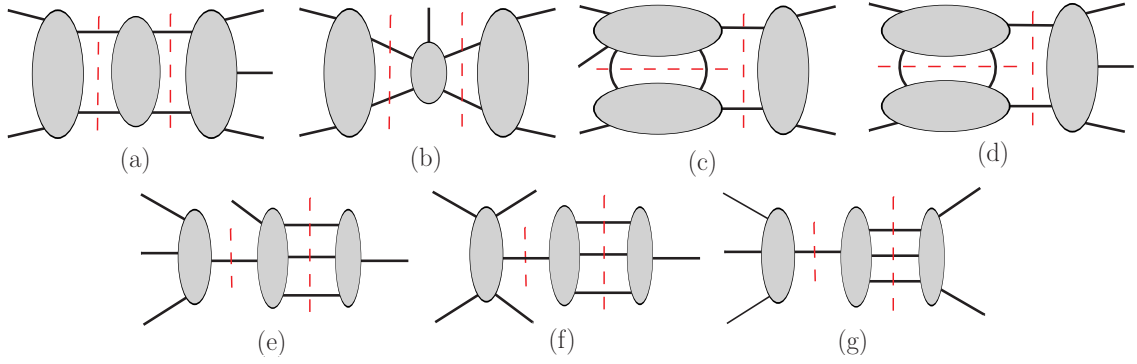}}
\caption[a]{\small The four-dimensional cuts used to verify the duality-satisfying form of the two-loop five-point $\NeqFour$~sYM amplitude. All cyclically distinct leg orderings on each blob are counted. Cuts (a) and (d) are also evaluated through a $D$-dimensional boxcut.
}
\label{mhvCutsFigure}
\end{figure}

The above two-loop $\NeqFour$ sYM amplitude has been verified using unitarity cuts in both general dimension $D$ and in $D=4$. The four-dimensional cuts are displayed in~\fig{mhvCutsFigure}; cuts (a)-(d) have been evaluated using the methods of ref.~\cite{Supersums}. The cuts (e)-(g) vanish because of $\NeqFour$ supersymmetry, as is consistent with the amplitudes given above. Together these cuts detect most of the possible terms that can show up in a generic five-point two-loop amplitude. At the very least, they detect all the numerator terms in the graphs of \fig{2L1}  that are at most quadratic in the loop momenta, implying that any potentially missing four-dimensional contributions would have to individually violate the expected UV power counting bound~\cite{BernDixonUV,HoweStelle,BRY}.

In addition, we have performed all nontrivial $D$-dimensional two-particle cuts that split the amplitude into a one-loop four-point amplitude times a five-point tree amplitude; these are easily calculated through the boxcut method described in~\cite{Neq44np}. 
For the above two-loop $\NeqEight$ supergravity amplitude one may compute all $D$-dimensional cuts using the input from the two-loop $\NeqFour$ sYM amplitude, using the method prescribed in refs.~\cite{compact3,superfiniteness,Gravityfourloops,Supersums}. However, the double-copy form of a duality-satisfying amplitude will automatically satisfy cuts that are evaluated this way. We will defer further verification of the two-loop amplitudes to future work.

\subsection{UV divergences at two loops}

\begin{figure}[t]
\begin{center}
\includegraphics[width=0.42 \textwidth]{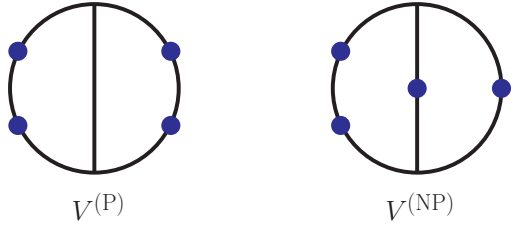}
\end{center}
\caption{\label{PNPvacuums}\small The two vacuum integrals that capture the $D=7$ ultraviolet divergence for planar and nonplanar diagrams, respectively.}
\end{figure}

Given \eqn{TwoLoopSYMAmplitude} and \eqn{TwoLoopSGAmplitude} we can easily compute the logarithmic ultraviolet divergence that first occur in $D=7$ for both theories. Indeed, as is manifest in the calculated amplitudes, for $D<7$ both theories are finite at two loops. This is consistent with the behavior of the known four-point amplitudes~\cite{MarcusSagnottii,BernDixonUV}. For $\NeqFour$ sYM the $D=7-2\eps$ divergence arises from the ``double-box'' diagrams (d), (e) and (f) of \fig{2L1}, which can be expressed in terms of planar and nonplanar vacuum integrals shown in \fig{PNPvacuums}. We have
\bea
{\cal I}^{(\rm d)}\Bigl|_{\rm UV~pole}&=&-\frac{1}{s_{12}}N^{(\rm d)} C^{(\rm d)} V^{(\rm P)}\,,~~~~{\cal I}^{(\rm e)}\Bigl|_{\rm UV~pole}=-\frac{1}{s_{12}}N^{(\rm e)} C^{(\rm e)} V^{(\rm NP)}\,, \nn \\
{\cal I}^{(\rm f)}\Bigl|_{\rm UV~pole}&=&-\frac{1}{s_{12}}N^{(\rm f)} C^{(\rm f)} V^{(\rm NP)} \,,
\label{vred}
\eea
where the vacuum integrals are given by~\cite{BernDixonUV,Neq44np}
\be
V^{(\rm P)}=-\frac{\pi}{20(4\pi)^7\eps}\,,~~~~~~~~
V^{(\rm NP)}=-\frac{\pi}{30(4\pi)^7\eps}\,.
\ee
Plugging in the vacuum diagram reduction~(\ref{vred}) into \eqn{TwoLoopSYMAmplitude}, and assuming a gauge group $SU(N_c)$, gives the divergence
\bea
{\cal A}_5^{(2)}\Bigr|_{\rm UV} &=& 
 -g^{7}\Bigl[\Bigl( N_c^2 V^{(\rm P)} + 12 (V^{(\rm P)}+V^{(\rm NP)})\Bigr)\Tr_{12345}\Bigl( 5 \beta_{12345} + \frac{\gamma_{12}}{s_{12}} (s_{35} - 2 s_{12}) \nn \\ &&
~~~~~~~\null+ \frac{\gamma_{23}}{s_{23}} ( s_{14} - 2 s_{23})+ \frac{\gamma_{34}}{s_{34}} (s_{25} - 2 s_{34}) + \frac{\gamma_{45}}{s_{45}} (s_{13} - 2 s_{45})+ \frac{\gamma_{51}}{s_{15}} ( s_{24} - 2 s_{15}) \Bigr)  \nn \\ &&
~~~~~~~\null-12 N_c (V^{(\rm P)} + V^{(\rm NP)}) \Tr_{123}\Tr_{45}\, s_{45} \Bigl(\frac{\gamma_{12}}{s_{12}}+\frac{\gamma_{23}}{s_{23}}+\frac{\gamma_{31}}{s_{13}}\Bigr)+{\rm perms}\Bigl]\,,
\label{TwoLoopSYMDiv}
\eea
where $\Tr_{12\cdots n}=\Tr(T^{a_1}T^{a_2} \cdots T^{a_n})$ encodes the gauge-group trace structures. The remaining trace structures, hidden in the ``$+$ perms,'' can be obtained from the given ones by using crossing symmetry.

%
%

For the $\NeqEight$ supergravity divergence, we have divergent contributions coming from all diagrams (a) through (f), nonetheless the structure of the divergence is similar to the  $\NeqFour$ one. The $D=7-2\eps$ supergravity divergence is
\be
{\cal M}_5^{(2)}\Bigr|_{\rm UV} = 
i \left(\frac{\kappa}{2}\right)^{7} \, 
\frac{1}{6} (V^{(\rm P)}+V^{(\rm NP)}) \sum_{S_5} \,  \frac{\gamma_{12}^2}{s_{12}} (s_{34}^2 + s_{35}^2 + s_{45}^2 - 3 s_{12}^2)\,,
\label{TwoLoopSGDiv}
\ee
where a 120-fold sum over permutations $S_5$ is left unevaluated. As before, we defined $\gamma_{ij}^2\equiv \gamma_{ij} {\tilde \gamma}_{ij}=  \gamma_{ij} (\gamma_{ij} |_{\eta_i^A \rightarrow \eta_i^{A+4}})$. 

For this calculation, there are two handy nontrivial relations that can be used to obtain the simple form (\ref{TwoLoopSGDiv}), namely
\bea
0&=& \sum_{S_5} \,  s_{12} (\gamma_{13} \gamma_{23} + \gamma_{14} \gamma_{24} + \gamma_{15} \gamma_{25} - 2 \gamma_{12}^2)\,, \nn \\ 
0&=& \sum_{S_5} \,  s_{12} (\gamma_{43} \gamma_{35} + \gamma_{34} \gamma_{45} + \gamma_{35} \gamma_{54})\,.
\eea

As at one loop, the forms of the five-point two-loop divergences are compatible with the logarithmic divergences observed for the four-point  amplitudes of the same theories~\cite{MarcusSagnottii,BernDixonUV}. Indeed, the precise four-point divergences are recovered in the factorization channels $s_{ij}\rightarrow0$. Although there are some local terms present in (\ref{TwoLoopSYMDiv}) and (\ref{TwoLoopSGDiv}), they seem closely tied to the terms with poles, suggesting that the former terms are simply the gauge-invariant completions of the singular terms. It is therefore likely that the counterterms at five points remain the same as at four points; namely, they are of the schematic forms $\partial^2 F^4$ and $\partial^4 R^4$ for gluon and graviton components, respectively. A direct calculation of the counterterm contributions to the five-point amplitudes would resolve any doubt.

\section{Conclusions}
\label{Conclusions}

In this paper we have presented the one- and two-loop five-point amplitudes of $\NeqFour$ super-Yang-Mills theory and $\NeqEight$ supergravity, valid for any value of the dimensional regularization parameter $D$, and for any Yang-Mills gauge group. The amplitudes are given in a representation that satisfies the duality between color and kinematic structures of each individual integral diagram. These amplitudes have been verified using a combination of four-dimensional and $D$-dimensional unitarity cuts. This shows that the duality is present for these particular amplitudes of the two theories, and strongly suggests that other multiloop and multileg amplitudes of these theories should similarly obey the duality. In the accompanying paper~\cite{threeloopfivept}, the duality-satisfying three-loop five-point amplitudes of the same two theories are worked out using methods identical to those presented here, thus adding further evidence in favor of the conjecture.

As can be expected from the duality between color and kinematics, the $\NeqEight$ supergravity amplitudes presented here have the property that individual diagram numerators are double copies of the corresponding ones of the $\NeqFour$ sYM theory. This provides further evidence for the claims that gravity theories are simply double copies of gauge theories, order by order, in perturbation theory~\cite{BCJ, BCJLoop}. To see this structural simplicity of gravity one must treat the kinematic structures on equal footing with gauge-group color structures, as done for the presented amplitudes. 

In the course of this work we have clarified the general structure of duality-satisfying five-point amplitudes in $\NeqFour$ sYM theory and $\NeqEight$ supergravity. The $\NeqFour$ sYM amplitudes have a natural decomposition in terms of six independent nonlocal kinematic prefactors, the remaining factors entering the diagram numerators are strictly local. 
A natural question is if higher-point amplitudes offer a similar decomposition; an investigation of the six-point one- and two-loop amplitudes would provide an excellent testing ground. Also of interest would be to explore one-loop five- and six-point amplitudes where the external momenta are in $D>4$ dimensions; knowing the explicit $D$-dimensional duality-satisfying forms will be helpful for understanding the constraints imposed by the duality. 

Although the duality at loop level has been observed for the two-loop four-point all-plus-helicity QCD amplitude, much of the evidence derives from the maximally supersymmetric theories. Arriving at duality-satisfying amplitudes for less-than-maximal supersymmetric theories is therefore of vital importance for the conjecture. One-loop amplitudes are by now well-studied in many theories, thus one would expect that the task of finding many interesting one-loop examples supporting the duality should be within reach. 

Exploring the duality at higher-loop levels should also be readily accomplishable. Indeed in parallel work with Bern, Dixon and Roiban we demonstrate that the duality between color and kinematics can easily be established at four loops for the $\NeqFour$ sYM  and $\NeqEight$ supergravity theories. Continuing this to even higher loops should make it possible to further address the question of the ultraviolet behavior of $\NeqEight$ supergravity.


\section*{Acknowledgments}
\vskip -.3 cm
We thank  Zvi Bern,  Camille Boucher-Veronneau, Johannes Br\"{o}del, Tristan Dennen, Lance Dixon, Daniel Freedman, Yu-tin Huang, Harald Ita, Renata Kallosh, Gregory \mbox{Korchemsky}, David Kosower and Radu Roiban for stimulating discussions on this work and related subjects.  A portion of this work was
completed at the Kavli Institute for Theoretical Physics, which the authors thank warmly for its hospitality.  JJMC gratefully acknowledges the Stanford Institute for Theoretical Physics for financial support. HJ's research is supported by the European Research Council under Advanced Investigator Grant ERC-AdG-228301.  This research was supported in part by the National Science Foundation under Grant No. NSF PHY05-51164. The figures were generated using Jaxodraw~\cite{Jaxo1and2}, based on Axodraw~\cite{Axo}.


\vskip1cm

\begin{thebibliography}{99}

\bibitem{NeqFourDefinition}
  L.~Brink, J.~H.~Schwarz and J.~Scherk,
  Nucl.\ Phys.\ B {\bf 121}, 77 (1977);\\
  F.~Gliozzi, J.~Scherk and D.~I.~Olive,
  Nucl.\ Phys.\ B {\bf 122}, 253 (1977).


\bibitem{MagicIdentities}
J.~M.~Drummond, J.~Henn, V.~A.~Smirnov and E.~Sokatchev,
JHEP {\bf 0701}, 064 (2007)
[hep-th/0607160];\\
%
  Z.~Bern, M.~Czakon, L.~J.~Dixon, D.~A.~Kosower and V.~A.~Smirnov,
  Phys.\ Rev.\  D {\bf 75}, 085010 (2007)
  [arXiv:hep-th/0610248].


\bibitem{DualSuperconformalSymmetry}
J.~M.~Drummond, J.~Henn, G.~P.~Korchemsky and E.~Sokatchev,
Nucl.\ Phys.\  B {\bf 828}, 317 (2010)
[arXiv:0807.1095 [hep-th]];\\
%
A.~Brandhuber, P.~Heslop and G.~Travaglini,
Phys.\ Rev.\  D {\bf 78}, 125005 (2008)
[arXiv:0807.4097 [hep-th]].

\bibitem{WittenTwistorString}
  E.~Witten,
  Commun.\ Math.\ Phys.\  {\bf 252}, 189 (2004)
  [arXiv:hep-th/0312171].

\bibitem{Grassmannians}
  N.~Arkani-Hamed, F.~Cachazo, C.~Cheung and J.~Kaplan,
  JHEP {\bf 1003}, 020 (2010)
  [arXiv:0907.5418 [hep-th]];\\
  %
  L.~J.~Mason and D.~Skinner,
  JHEP {\bf 0911}, 045 (2009)
  [arXiv:0909.0250 [hep-th]];\\
  %
  N.~Arkani-Hamed, F.~Cachazo and C.~Cheung,
  JHEP {\bf 1003}, 036 (2010)
  [arXiv:0909.0483 [hep-th]].



\bibitem{BCJ}
Z.~Bern, J.~J.~M.~Carrasco and H.~Johansson,
Phys.\ Rev.\  D {\bf 78}, 085011 (2008)
[arXiv:0805.3993 [hep-ph]].

\bibitem{BCJLoop}
Z.~Bern, J.~J.~M.~Carrasco and H.~Johansson,
Phys.\ Rev.\ Lett.\  {\bf 105}, 061602 (2010)
[arXiv:1004.0476 [hep-th]].

\bibitem{Monodromy}
  N.~E.~J.~Bjerrum-Bohr, P.~H.~Damgaard and P.~Vanhove,
  Phys.\ Rev.\ Lett.\  {\bf 103}, 161602 (2009)
  [arXiv:0907.1425 [hep-th]];
  %
  S.~Stieberger,
  arXiv:0907.2211 [hep-th].

\bibitem{amplituderelationProof}
  B.~Feng, R.~Huang and Y.~Jia,
  Phys.\ Lett.\  B {\bf 695}, 350 (2011)
  [arXiv:1004.3417 [hep-th]];\\
  %
  Y.~X.~Chen, Y.~J.~Du and B.~Feng,
  JHEP {\bf 1102}, 112 (2011)
  [arXiv:1101.0009 [hep-th]].

\bibitem{ck4l}
 Z.~Bern, J.~J.~M.~Carrasco, L.~J.~Dixon, H.~Johansson and R.~Roiban, 
 to appear.

\bibitem{Square}
Z.~Bern, T.~Dennen, Y.~t.~Huang and M.~Kiermaier,
Phys.\ Rev.\  D {\bf 82}, 065003 (2010)
[arXiv:1004.0693 [hep-th]].

\bibitem{CremmerJulia}
  E.~Cremmer, B.~Julia and J.~Scherk,
  Phys.\ Lett.\ B {\bf 76}, 409 (1978);\\
E.~Cremmer and B.~Julia,
Phys.\ Lett.\ B {\bf 80}, 48 (1978);
Nucl.\ Phys.\  B {\bf 159}, 141 (1979).

\bibitem{KLT}
H.~Kawai, D.~C.~Lewellen and S.~H.~H.~Tye,
Nucl.\ Phys.\ B {\bf 269}, 1 (1986);
%
Z.~Bern,
Living Rev.\ Rel.\  {\bf 5}, 5 (2002)
[arXiv:gr-qc/0206071].


\bibitem{Bjerrum2}
  N.~E.~J.~Bjerrum-Bohr, P.~H.~Damgaard, T.~Sondergaard and P.~Vanhove,
  JHEP {\bf 1006}, 003 (2010)
  [arXiv:1003.2403 [hep-th]];\\
%
  C.~R.~Mafra,
  JHEP {\bf 1011}, 096 (2010)
  [arXiv:1007.3639 [hep-th]].

\bibitem{Tye}
S.~H.~Henry Tye and Y.~Zhang,
JHEP {\bf 1006}, 071 (2010)
[arXiv:1003.1732 [hep-th]].

\bibitem{KiermaierTalk}
M.~Kiermaier, Amplitudes 2010, Queen Mary, University of London,\\
{\tt http://www.strings.ph.qmul.ac.uk/$\sim$theory/Amplitudes2010/Talks/MK2010.pdf}

\bibitem{BBDSVsolution}
  N.~E.~J.~Bjerrum-Bohr, P.~H.~Damgaard, T.~Sondergaard and P.~Vanhove,
  JHEP {\bf 1101}, 001 (2011)
  [arXiv:1010.3933 [hep-th]].
%

\bibitem{Feng}
  B.~Feng, R.~Huang and Y.~Jia,
  Phys.\ Lett.\  B {\bf 695}, 350 (2011)
  [arXiv:1004.3417 [hep-th]].

\bibitem{ConnellKinematicAlgebra}
  R.~Monteiro and D.~O'Connell,
  JHEP {\bf 1107}, 007 (2011)
  [arXiv:1105.2565 [hep-th]].


\bibitem{StiebergerNumerators}
  C.~R.~Mafra, O.~Schlotterer and S.~Stieberger,
  JHEP {\bf 1107}, 092 (2011)
  [arXiv:1104.5224 [hep-th]].

\bibitem{BCJOther}
  T.~Sondergaard,
  Nucl.\ Phys.\  B {\bf 821}, 417 (2009)
  [arXiv:0903.5453 [hep-th]];\\
  %
C.~R.~Mafra,
JHEP {\bf 1001}, 007 (2010)
[arXiv:0909.5206 [hep-th]];\\
%
  P.~Vanhove,
  arXiv:1004.1392 [hep-th];\\
%
  Y.~Jia, R.~Huang and C.~Y.~Liu,
  Phys.\ Rev.\  D {\bf 82}, 065001 (2010)
  [arXiv:1005.1821 [hep-th]];\\	
%
  N.~E.~J.~Bjerrum-Bohr, P.~H.~Damgaard, B.~Feng and T.~Sondergaard,
  Phys.\ Rev.\  D {\bf 82}, 107702 (2010)
  [arXiv:1005.4367 [hep-th]];
%
  Phys.\ Lett.\  B {\bf 691}, 268 (2010)
  [arXiv:1006.3214 [hep-th]];
%
  B.~Feng and S.~He,
  JHEP {\bf 1009}, 043 (2010)
  [arXiv:1007.0055 [hep-th]];\\
%
  N.~E.~J.~Bjerrum-Bohr, P.~H.~Damgaard, B.~Feng and T.~Sondergaard,
  JHEP {\bf 1009}, 067 (2010)
  [arXiv:1007.3111 [hep-th]];\\
%
  D.~Vaman and Y.~P.~Yao,
  JHEP {\bf 1011}, 028 (2010)
  [arXiv:1007.3475 [hep-th]];\\
%
  C.~R.~Mafra,
  arXiv:1007.4999 [hep-th];\\
%
  B.~Feng, S.~He, R.~Huang and Y.~Jia,
  JHEP {\bf 1010}, 109 (2010)
  [arXiv:1008.1626 [hep-th]];\\
%
  Y.~Abe,
  Nucl.\ Phys.\  B {\bf 842}, 475 (2011)
  [arXiv:1008.2800 [hep-th]];\\
%
  J.~Bjornsson,
  JHEP {\bf 1101}, 002 (2011)
  [arXiv:1009.5906 [hep-th]];\\
  %
  J.~H.~Huang, R.~Huang and Y.~Jia,
  J.\ Phys.\ A  {\bf 44}, 425401 (2011)
  [arXiv:1009.5073 [hep-th]];\\
%
  C.~R.~Mafra, O.~Schlotterer, S.~Stieberger and D.~Tsimpis,
  Nucl.\ Phys.\  B {\bf 846}, 359 (2011)
  [arXiv:1011.0994 [hep-th]];\\
  %
  H.~Nastase and H.~J.~Schnitzer,
  JHEP {\bf 1101}, 048 (2011)
  [arXiv:1011.2487 [hep-th]];\\
  %
  J.~Broedel and R.~Kallosh,
  JHEP {\bf 1106}, 024 (2011)
  [arXiv:1103.0322 [hep-th]];\\
%
  N.~E.~J.~Bjerrum-Bohr, P.~H.~Damgaard, H.~Johansson and T.~Sondergaard,
  JHEP {\bf 1105}, 039 (2011)
  [arXiv:1103.6190 [hep-th]];\\
  %
  Y.~J.~Du, B.~Feng and C.~H.~Fu,
  JHEP {\bf 1108}, 129 (2011)
  [arXiv:1105.3503 [hep-th]];\\
  %
  C.~R.~Mafra, O.~Schlotterer and S.~Stieberger,
  arXiv:1106.2645 [hep-th].


\bibitem{threeloopfivept}
 J.~J.~M.~Carrasco and H.~Johansson, to appear.

\bibitem{BernDennenTrace}
  Z.~Bern and T.~Dennen,
  Phys.\ Rev.\ Lett.\  {\bf 107}, 081601 (2011)
  [arXiv:1103.0312 [hep-th]].

\bibitem{MHVoneloop}
  Z.~Bern, L.~J.~Dixon, D.~C.~Dunbar and D.~A.~Kosower,
  Nucl.\ Phys.\  B {\bf 425}, 217 (1994)
  [arXiv:hep-ph/9403226].
  

\bibitem{KorchemskyOneLoop}
  J.~M.~Drummond, J.~Henn, G.~P.~Korchemsky and E.~Sokatchev,
  arXiv:0808.0491 [hep-th].

\bibitem{FivePtBDS}
  Z.~Bern, M.~Czakon, D.~A.~Kosower, R.~Roiban and V.~A.~Smirnov,
  Phys.\ Rev.\ Lett.\  {\bf 97}, 181601 (2006)
  [arXiv:hep-th/0604074].

\bibitem{CachazoLeadingSingularityAndCalcs}
  F.~Cachazo,
  arXiv:0803.1988 [hep-th].

\bibitem{Spradlin3loop}
  M.~Spradlin, A.~Volovich and C.~Wen,
  Phys.\ Rev.\  D {\bf 78}, 085025 (2008)
  [arXiv:0808.1054 [hep-th]].

\bibitem{BRY}
Z.~Bern, J.~S.~Rozowsky and B.~Yan,
Phys.\ Lett.\  B {\bf 401}, 273 (1997)
[hep-ph/9702424].


\bibitem{compact3}
  Z.~Bern, J.~J.~M.~Carrasco, L.~J.~Dixon, H.~Johansson and R.~Roiban,
  Phys.\ Rev.\  D {\bf 78}, 105019 (2008)
  [arXiv:0808.4112 [hep-th]].

\bibitem{Neq44np}
  Z.~Bern, J.~J.~M.~Carrasco, L.~J.~Dixon, H.~Johansson and R.~Roiban,
  Phys.\ Rev.\  D {\bf 82}, 125040 (2010)
  [arXiv:1008.3327 [hep-th]].
  
\bibitem{MHVOneLoopGravity}
Z.~Bern, L.~J.~Dixon, M.~Perelstein and J.~S.~Rozowsky,
Nucl.\ Phys.\ B {\bf 546}, 423 (1999)
[hep-th/9811140].


\bibitem{UVreview}
 Z.~Bern, J.~J.~Carrasco, L.~J.~Dixon, H.~Johansson and R.~Roiban,
 Fortschr.\ Phys.\ {\bf 59}, 561 (2011)
 [arXiv:1103.1848 [hep-th]];\\
%
 L.~J.~Dixon,
 arXiv:1005.2703 [hep-th];\\
 %
 Z.~Bern, J.~J.~M.~Carrasco and H.~Johansson,
 arXiv:0902.3765 [hep-th].

\bibitem{StringsE7andSusy}
  M.~B.~Green, J.~G.~Russo, P.~Vanhove,
  JHEP {\bf 1006}, 075 (2010).
  [arXiv:1002.3805 [hep-th]].
  %

 \bibitem{BeisertCountertermsandElvangReview}
  H.~Elvang, D.~Z.~Freedman and M.~Kiermaier,
  J.\ Phys.\ A  {\bf 44}, 454009 (2011)
  [arXiv:1012.3401 [hep-th]];\\
%
 N.~Beisert, H.~Elvang, D.~Z.~Freedman, M.~Kiermaier, A.~Morales and S.~Stieberger,
 Phys.\ Lett.\  B {\bf 694}, 265 (2010)
 [arXiv:1009.1643 [hep-th]];\\
 %
  H.~Elvang, D.~Z.~Freedman and M.~Kiermaier,
  JHEP {\bf 1011}, 016 (2010)
  [arXiv:1003.5018 [hep-th]].

 \bibitem{RecentDualityStuff}
  %
  G.~Bossard, C.~Hillmann, H.~Nicolai,
  JHEP {\bf 1012}, 052 (2010).
  [arXiv:1007.5472 [hep-th]];\\
%
  R.~Kallosh,
   [arXiv:1103.4115 [hep-th]];\\
%
  G.~Bossard and H.~Nicolai,
  JHEP {\bf 1108}, 074 (2011)
  [arXiv:1105.1273 [hep-th]].
  

\bibitem{HoweStelle}
  P.~S.~Howe and K.~S.~Stelle,
  Phys.\ Lett.\  B {\bf 554}, 190 (2003)
  [arXiv:hep-th/0211279].

\bibitem{issupergravityfinite}
  Z.~Bern, L.~J.~Dixon and R.~Roiban,
  Phys.\ Lett.\  B {\bf 644}, 265 (2007)
  [arXiv:hep-th/0611086].


\bibitem{superfiniteness}
  Z.~Bern, J.~J.~Carrasco, L.~J.~Dixon, H.~Johansson, D.~A.~Kosower and R.~Roiban,
  Phys.\ Rev.\ Lett.\  {\bf 98}, 161303 (2007)
  [arXiv:hep-th/0702112].

\bibitem{Gravityfourloops}
  Z.~Bern, J.~J.~Carrasco, L.~J.~Dixon, H.~Johansson and R.~Roiban,
  Phys.\ Rev.\ Lett.\  {\bf 103}, 081301 (2009)
  [arXiv:0905.2326 [hep-th]].


\bibitem{Bjornsson}
  J.~Bjornsson and M.~B.~Green,
  JHEP {\bf 1008}, 132 (2010)
  [arXiv:1004.2692 [hep-th]].

\bibitem{BossardVanishingVolume}
  G.~Bossard, P.~S.~Howe, K.~S.~Stelle and P.~Vanhove,
  Class.\ Quant.\ Grav.\  {\bf 28}, 215005 (2011)
  [arXiv:1105.6087 [hep-th]];\\
  %
  G.~Bossard, P.~S.~Howe and K.~S.~Stelle,
  JHEP {\bf 1101}, 020 (2011)
  [arXiv:1009.0743 [hep-th]].
  

\bibitem{Kallosh7loops}
  R.~Kallosh, P.~Ramond,
  [arXiv:1006.4684 [hep-th]];\\
  R.~Kallosh,
  Phys.\ Rev.\  D {\bf 80}, 105022 (2009)
  [arXiv:0903.4630 [hep-th]].


\bibitem{GSB}
  M.~B.~Green, J.~H.~Schwarz and L.~Brink,
  Nucl.\ Phys.\  B {\bf 198}, 474 (1982).
  
\bibitem{MarcusSagnottii}
  N.~Marcus and A.~Sagnotti,
  Nucl.\ Phys.\  B {\bf 256}, 77 (1985).

\bibitem{BernDixonUV}
  Z.~Bern, L.~J.~Dixon, D.~C.~Dunbar, M.~Perelstein and J.~S.~Rozowsky,
  Nucl.\ Phys.\  B {\bf 530}, 401 (1998)
  [arXiv:hep-th/9802162].


\bibitem{UnitarityMethod}
Z.~Bern, L.~J.~Dixon, D.~C.~Dunbar and D.~A.~Kosower,
Nucl.\ Phys.\ B {\bf 425}, 217 (1994)
[hep-ph/9403226];
%
Nucl.\ Phys.\ B {\bf 435}, 59 (1995)
[hep-ph/9409265].

\bibitem{GeneralizedUnitarity}
Z.~Bern, L.~J.~Dixon and D.~A.~Kosower,
Nucl.\ Phys.\ B {\bf 513}, 3 (1998)
[hep-ph/9708239].

\bibitem{GeneralizedUnitarity2}
Z.~Bern, V.~Del Duca, L.~J.~Dixon and D.~A.~Kosower,
Phys.\ Rev.\  D {\bf 71}, 045006 (2005)
[hep-th/0410224].

\bibitem{BCF}
R.~Britto, F.~Cachazo and B.~Feng,
Nucl.\ Phys.\  B {\bf 725}, 275 (2005)
[hep-th/0412103].

\bibitem{ZviYutinReview}
  Z.~Bern and Y.~t.~Huang,
  J.\ Phys.\ A  {\bf 44}, 454003 (2011)
  [arXiv:1103.1869 [hep-th]].

\bibitem{DixonReview}
  L.~J.~Dixon,
  J.\ Phys.\ A  {\bf 44}, 454001 (2011)
  [arXiv:1105.0771 [hep-th]].

\bibitem{JJMCHJreview}
  J.~J.~M.~Carrasco and H.~Johansson,
  J.\ Phys.\ A  {\bf 44}, 454004 (2011)
  [arXiv:1103.3298 [hep-th]].
  

\bibitem{BrittoReview}
  R.~Britto,
  J.\ Phys.\ A  {\bf 44}, 454006 (2011)
  [arXiv:1012.4493 [hep-th]].

 \bibitem{StructureReview}
  Z.~Bern, J.~J.~M.~Carrasco and H.~Johansson,
  Nucl.\ Phys.\ Proc.\ Suppl.\  {\bf 205-206}, 54 (2010)
  [arXiv:1007.4297 [hep-th]].

\bibitem{Supersums}
  Z.~Bern, J.~J.~M.~Carrasco, H.~Ita, H.~Johansson and R.~Roiban,
  Phys.\ Rev.\  D {\bf 80}, 065029 (2009)
  [arXiv:0903.5348 [hep-th]].

\bibitem{JJCMBroedel}
 J.~Br\"{o}del and J.~J.~M.~Carrasco,
  Phys.\ Rev.\  D {\bf 84}, 085009 (2011)
  [arXiv:1107.4802 [hep-th]].

\bibitem{conformal5}
Z.~Bern, J.~J.~M.~Carrasco, H.~Johansson and D.~A.~Kosower,
Phys.\ Rev.\  D {\bf 76}, 125020 (2007)
[arXiv:0705.1864 [hep-th]].

\bibitem{DimShift}
  Z.~Bern, L.~J.~Dixon, D.~C.~Dunbar and D.~A.~Kosower,
  Phys.\ Lett.\  B {\bf 394}, 105 (1997)
  [arXiv:hep-th/9611127].

\bibitem{BernMorgan}
  Z.~Bern and A.~G.~Morgan,
  Nucl.\ Phys.\  B {\bf 467}, 479 (1996)
  [arXiv:hep-ph/9511336].

\bibitem{Jaxo1and2}
D.~Binosi and L.~Theussl,
Comput.\ Phys.\ Commun.\ {\bf 161}, 76 (2004)
[hep-ph/0309015];\\
D.~Binosi, J.~Collins, C.~Kaufhold and L.~Theussl,
Comput.\ Phys.\ Commun.\  {\bf 180}, 1709 (2009)
[arXiv:0811.4113 [hep-ph]].

\bibitem{Axo}
J.~A.~M.~Vermaseren,
Comput.\ Phys.\ Commun.\ {\bf 83}, 45 (1994).



  


\end{thebibliography}
\end{document}